\documentclass[sigconf,authorversion,nonacm]{acmart}

\usepackage{booktabs} 
\usepackage{amsmath} 

\usepackage[ruled]{algorithm2e} 

\SetAlFnt{\small}
\SetAlCapFnt{\small}
\SetAlCapNameFnt{\small}
\SetAlCapHSkip{0pt}

\usepackage{caption}
\usepackage{subcaption}
\usepackage[group-separator={,}]{siunitx}

\usepackage[capitalize]{cleveref}
\crefname{section}{\S}{\S}
\Crefname{section}{Section}{Sections}
\Crefname{table}{Table}{Tables}

\copyrightyear{2023}
\acmYear{2023}
\setcopyright{acmlicensed}\acmConference[DIGIPRO '23]{The Digital Production Symposium}{August 5, 2023}{Los Angeles, CA, USA}
\acmBooktitle{The Digital Production Symposium (DIGIPRO '23), August 5, 2023, Los Angeles, CA, USA}
\acmPrice{15.00}
\acmDOI{10.1145/3603521.3604293}
\acmISBN{979-8-4007-0238-9/23/08}

\begin{document}
\title{Magenta Green Screen: \texorpdfstring{\\}{}Spectrally Multiplexed Alpha Matting with Deep Colorization}

\author{Dmitriy Smirnov}
\affiliation{%
  \institution{Netflix}
  \city{Los Gatos}
  \state{CA}
  \country{USA}
}

\author{Chloe LeGendre}
\affiliation{%
  \institution{Netflix Eyeline Studios}
  \city{Los Angeles}
  \state{CA}
  \country{USA}
}

\author{Xueming Yu}
\affiliation{%
  \institution{Netflix Eyeline Studios}
  \city{Los Angeles}
  \state{CA}
  \country{USA}
}

\author{Paul Debevec}
\affiliation{%
  \institution{Netflix Eyeline Studios}
  \city{Los Angeles}
  \state{CA}
  \country{USA}
}

\renewcommand\shortauthors{Smirnov et al.}

\begin{abstract}
We introduce \emph{Magenta Green Screen}, a novel machine learning--enabled matting technique for recording the color image of a foreground actor and a simultaneous high-quality alpha channel without requiring a special camera or manual keying techniques. We record the actor on a green background but light them with only red and blue foreground lighting. In this configuration, the green channel shows the actor silhouetted against a bright, even background, which can be used directly as a holdout matte, the inverse of the actor's \emph{alpha channel}. We then restore the green channel of the foreground using a machine learning colorization technique. We train the colorization model with an example sequence of the actor lit by white lighting, yielding convincing and temporally stable colorization results. We further show that time-multiplexing the lighting between Magenta Green Screen and Green Magenta Screen allows the technique to be practiced under what appears to be mostly normal lighting. We demonstrate that our technique yields high-quality compositing results when implemented on a modern LED virtual production stage. The alpha channel data obtainable with our technique can provide significantly higher quality training data for natural image matting algorithms to support future ML matting research.
\vspace{-1.5em}
\end{abstract}




\begin{teaserfigure}
\centering
 \begin{subfigure}[t]{0.33\columnwidth}
     \centering
     \includegraphics[width=\textwidth]{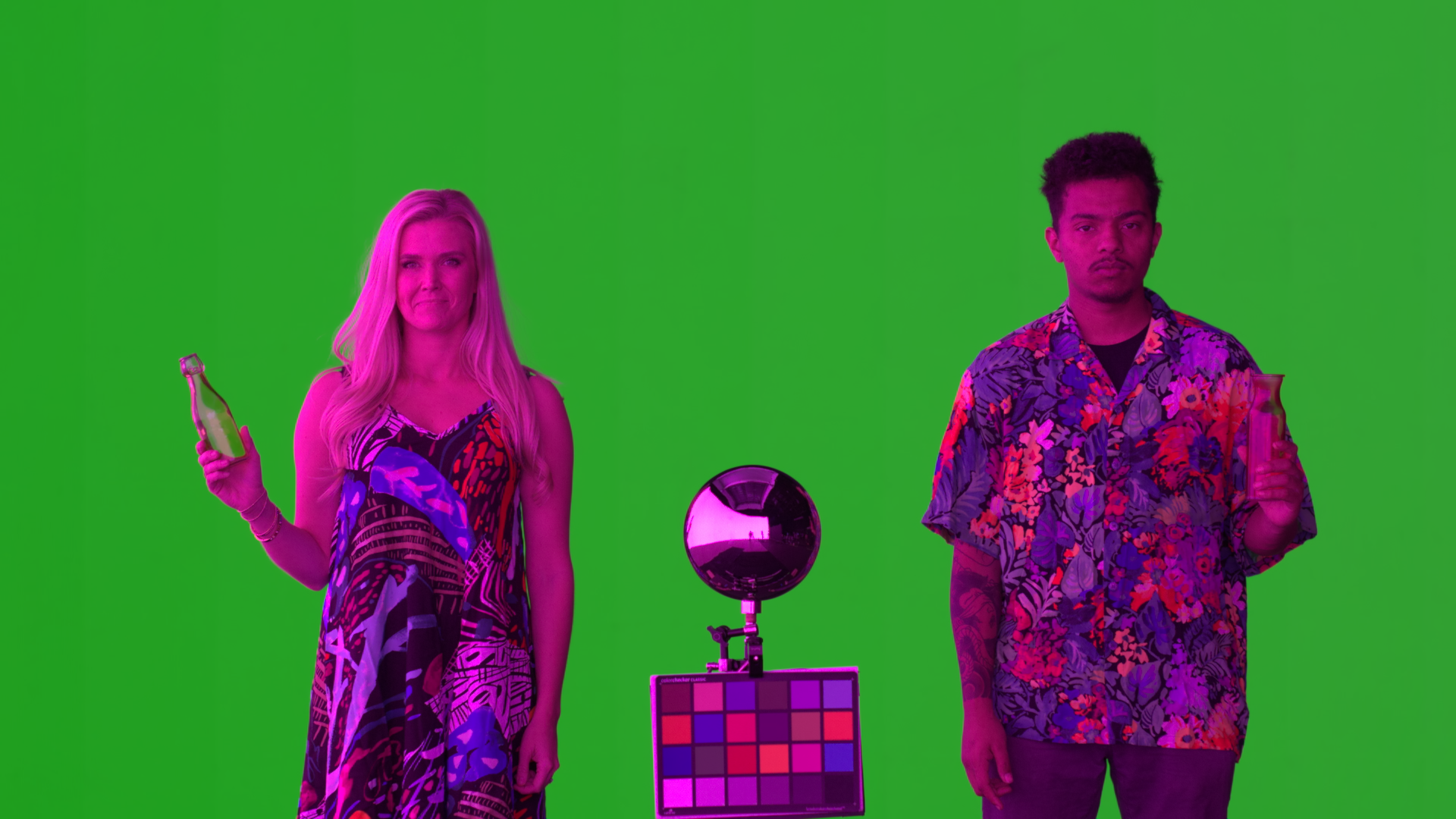}
     \caption{Magenta Green Screen}
     \label{fig:teaser-mg}
 \end{subfigure}
 \begin{subfigure}[t]{0.33\columnwidth}
     \centering
     \includegraphics[width=\textwidth]{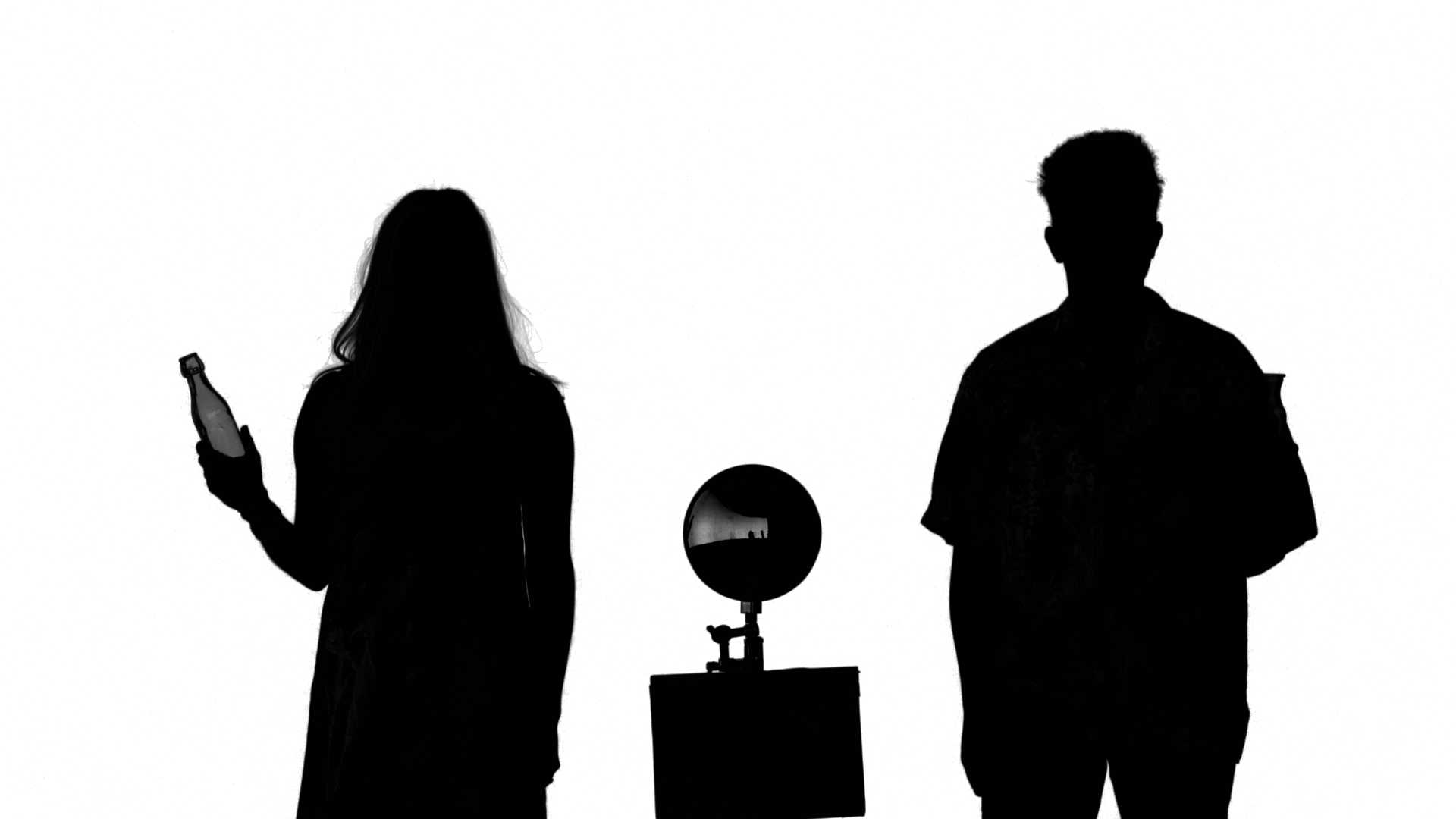}
     \caption{Derived Holdout Matte}
     \label{fig:teaser-alpha}
 \end{subfigure}
 \begin{subfigure}[t]{0.33\columnwidth}
     \centering
     \includegraphics[width=\textwidth]{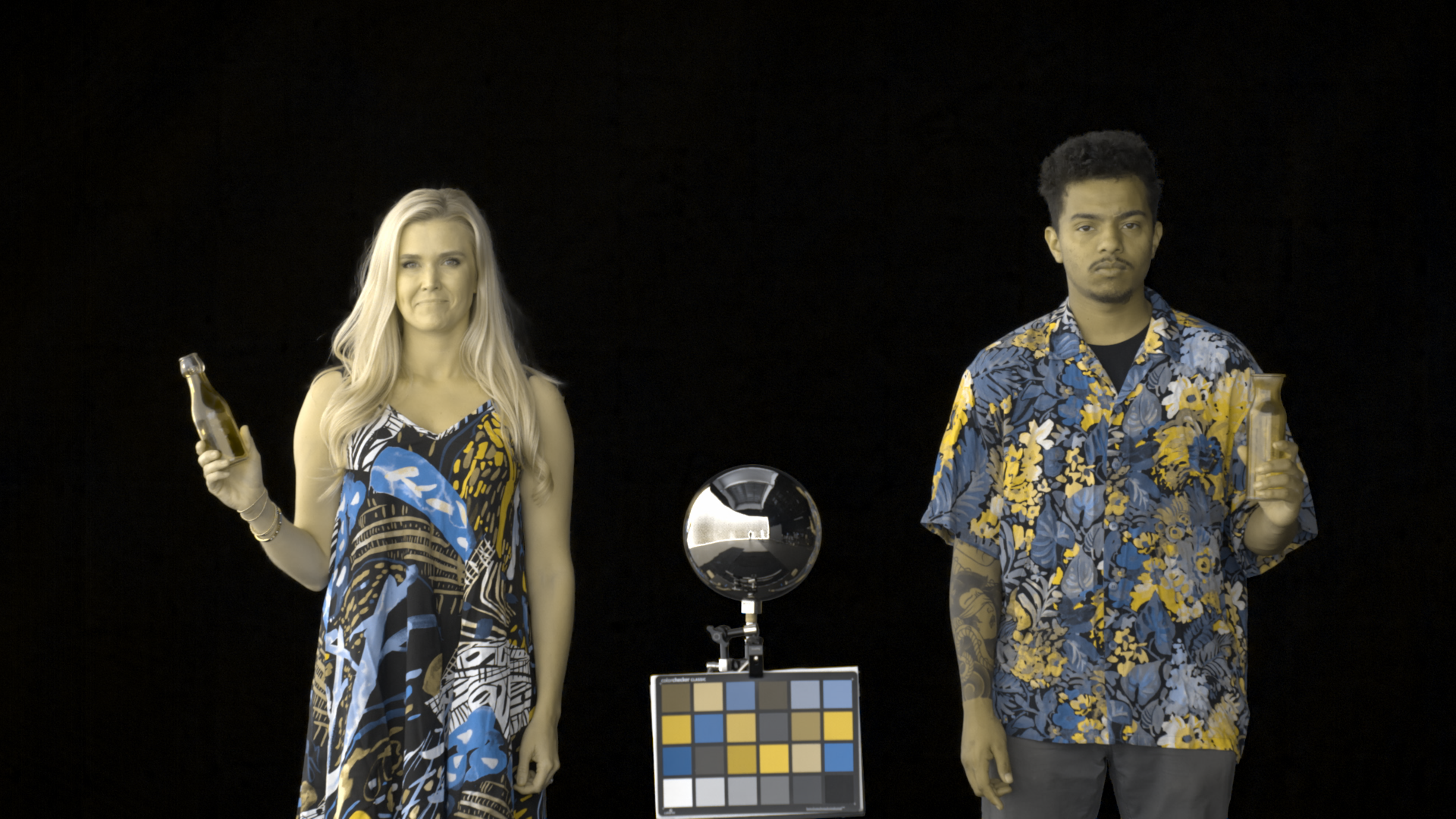}
     \caption{Naively Colorized Foreground}
     \label{fig:teaser-naive}
 \end{subfigure}\\
 \vspace{0.5em}
 \begin{subfigure}[t]{0.33\columnwidth}
     \centering
     \includegraphics[width=\textwidth]{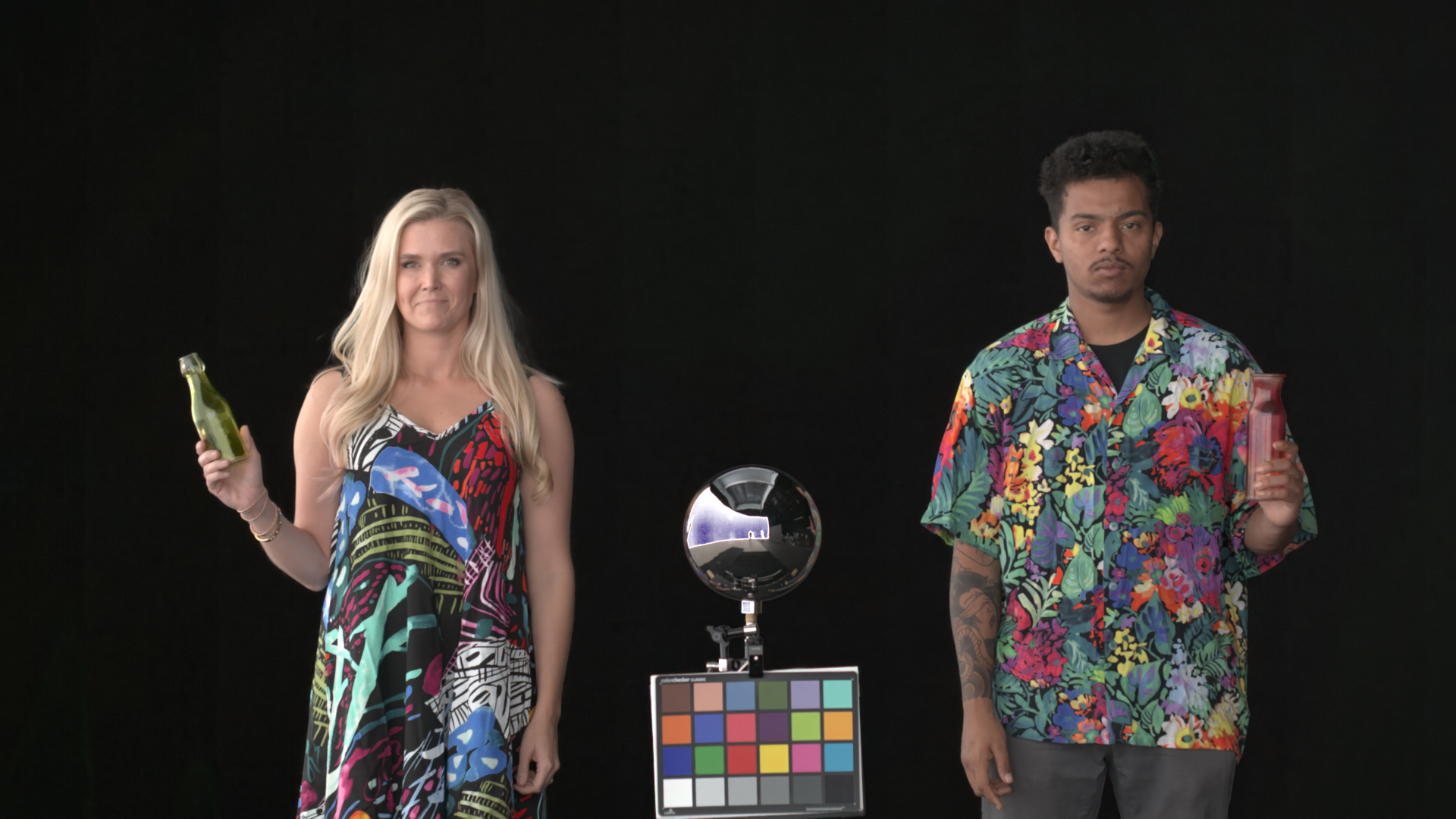}
     \caption{ML-Colorized Foreground}
     \label{fig:teaser-colorized}
 \end{subfigure}
 \begin{subfigure}[t]{0.33\columnwidth}
     \centering
     \includegraphics[width=\textwidth]{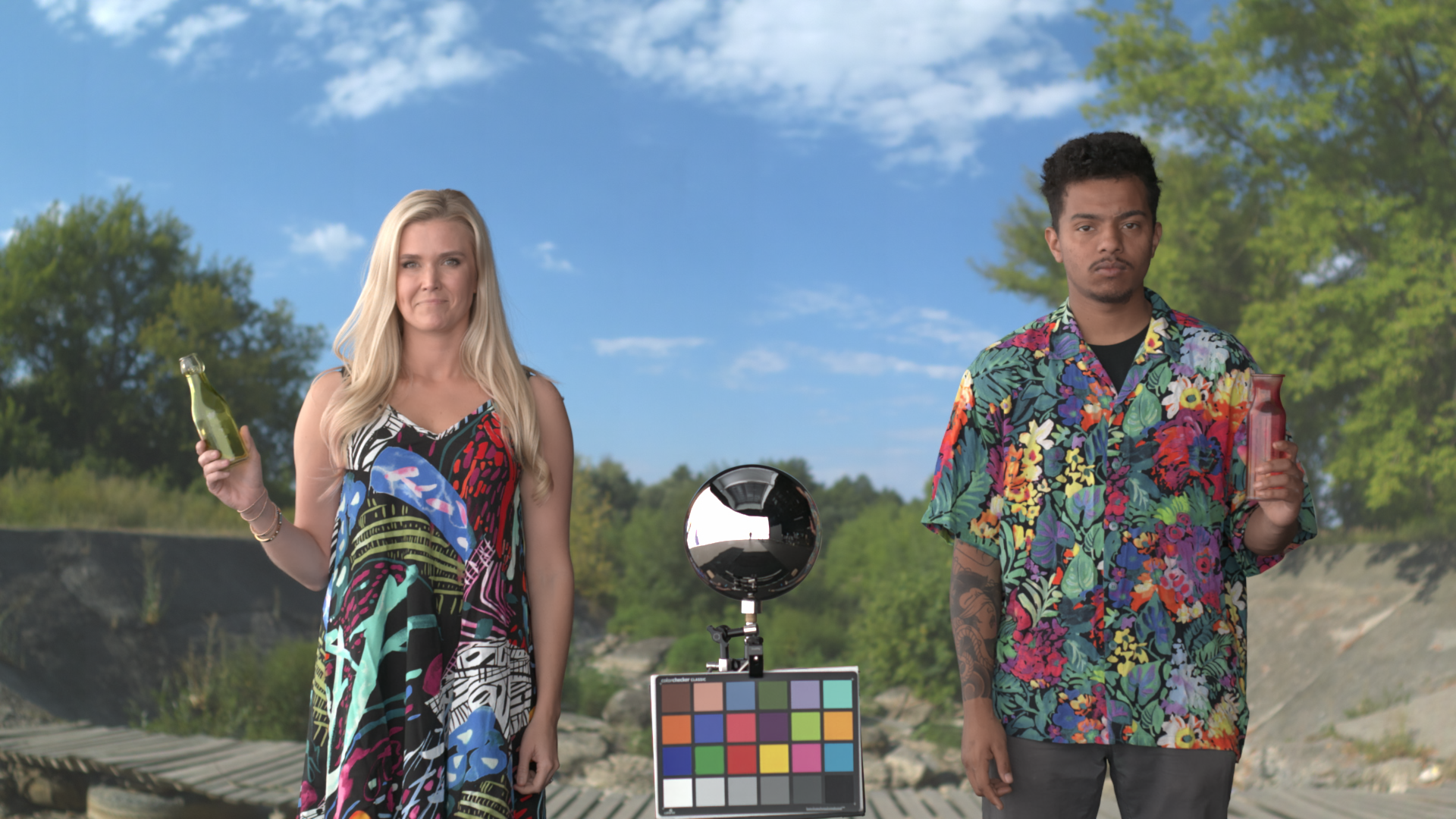}
     \caption{Composited onto Background}
     \label{fig:teaser-comp}
 \end{subfigure}
 \begin{subfigure}[t]{0.33\columnwidth}
     \centering
     \includegraphics[width=\textwidth]{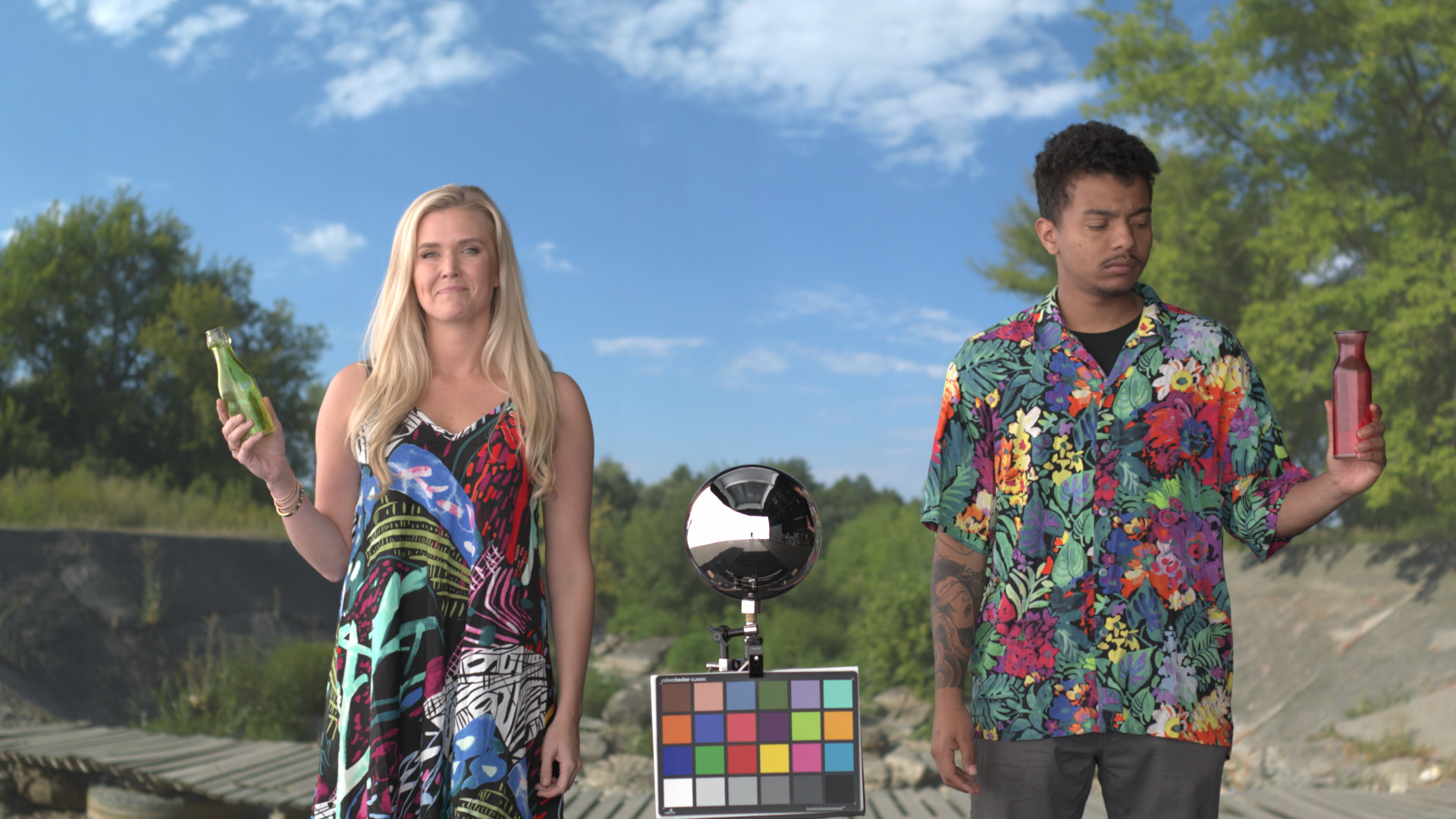}
     \caption{Ground Truth In-Camera Composite}
     \label{fig:teaser-gt}
 \end{subfigure}
 \vspace{-0.5em}
  \caption{We record actors on a virtual production LED stage using a green background and only red and blue foreground lighting, with no green foreground light illuminating the actors (a).  The processed green channel forms a high-quality holdout matte (b). We can naively restore the green channel of the foreground element as a linear combination of the red and blue foreground channels (c). For better accuracy, the green channel can instead be restored through an example-based AI technique (d). We composite the colorized foreground onto a background plate using the derived alpha (e). The result is virtually indistinguishable from the actors actually photographed in front of the background plate (f).\vspace{1em}}
  \label{fig:teaser}
\end{teaserfigure}

\maketitle

\section{Introduction}

Separating actors from a background image to composite them into a new scene is a fundamental problem in visual effects, and one which still poses challenges in the digital era \citep{Wright:2013:Digital}.  The problem is challenging since each pixel of an image can belong to both the foreground and the background: partial coverage at edges, wispy and transparent structures, defocused and motion-blurred areas all exhibit partial transparency.  Determining the RGB color of the foregound element at a pixel, as well as the pixel's transparency $\alpha$, is both tricky and underdetermined by a single RGB image.  Even when the actor is filmed in front of a green screen, it can be challenging to obtain a high-quality foreground element and $\alpha$ channel.  As stated by computer graphics pioneer Alvy Ray Smith, ``The history of digital image compositing is essentially the history of the \textit{alpha channel}.'' \citep{Smith:1995:Alpha}

Notably, while visual effects practitioners in both the film and digital cinematography eras have long relied on \textit{chroma-keying}, or filming an actor in front of a blue or green screen and then using color space manipulations to derive a foreground/background separation \citep{Sawicki:2007:Filming, Fielding:2013:Techniques, Vlahos:1964:Composite, Vlahos:1993:Traveling, Beyer:1965:Traveling}, these techniques rely on heuristics and approximations, without analytically solving the underconstrained matting equations, as detailed by \citet{Smith:1996:Blue}. As such, the contemporary green-screen keying algorithms in the modern compositor's toolkit, often part of proprietary commercial tools (Image-Based Keyer, Primatte, etc.), require substantial manual parameter tuning to work effectively and only provide an approximation of the per-pixel transparency \citep{Aksoy:2016:Interactive}.

To complement their analysis of the matting equations, \citet{Smith:1996:Blue} introduced \textit{triangulation matting}, whereby a stationary subject is filmed in front of two known backgrounds. Given this additional constraint---two known backgrounds instead of one---an accurate alpha channel can be derived from the matting equations. This technique was leveraged by \citet{Rhemann:2009:Perceptually} to develop the ground truth alpha channel imagery for the first public benchmark designed for fairly evaluating the myriad alpha matting algorithms proposed by the computer vision research community. Although limited to imagery of about thirty different relatively small, static objects, this benchmark\footnote{\url{http://www.alphamatting.com/datasets.php}} was the first of its kind to include analytically-derived ground truth alpha mattes recovered from photographs. 

Since the introduction of this benchmark, matting algorithms have received a great deal of attention from the visual computing research community, particularly algorithms aiming to solve the highly ill-posed \textit{natural image matting problem} where the background content is unknown and non-uniform \citep{Levin:2007:Closed}. Unfortunately, while several new public datasets and benchmarks have been released in the subsequent years, including those with imagery of people and those with motion imagery, the static and object-limited dataset of \citet{Rhemann:2009:Perceptually} remains, to the best of our knowledge, the only dataset which provides ground truth alpha mattes. The remaining datasets published in the era of deep learning--based matting algorithms (\citep{Lin:2021:Real, Erofeev:2015:Perceptually, Xu:2017:Deep}), leverage labels derived from chroma-key approximations or manual rotoscoping. This is an understandable limitation of these datasets, since methodologies used to record ground truth alpha mattes at video rates have remained difficult to practice, requiring sophisticated hardware configurations. We therefore introduce \textit{Magenta Green Screen} with the goal of recording ground truth alpha mattes at video rates, such that we can film scenes containing realistic human motion with appropriate motion blur along with diverse materials, including challenging-to-matte moving hair strands and transparent materials.

In our proposed technique, we use a three-channel RGB camera to record the red, blue and \textit{alpha} channels of the desired four-channel RGBA image, instead of the typical red, blue, and green channels. As our recorded imagery is thus missing its typical green channel, we infer this green channel given the corresponding red and blue channels using a machine learning network trained on full-color exemplar images.  We demonstrate that our proposed technique can be used directly as a film production methodology and argue that it can be used to capture realistic alpha channel data at scale to train deep learning--based video matting algorithms. While we record data to demonstrate our technique in an LED volume virtual production stage, this hardware setup is not strictly necessary: the only strict requirement is a set of LED based light sources capable of producing fields of red, green, and blue light independently.

\section{Related Work}

The process of deriving a matte (i.e., alpha channel) for foreground elements to composite them onto new backgrounds has a rich history with a long line of contributions from both industry and academia.  While a complete discussion is outside the scope of this paper, in this section we review some of the most relevant literature to our project.

Where possible, actors are filmed in front of a green screen, and both the actors' foreground appearance and alpha channel are derived from the green screen image.  The central difficulty is that while the green channel is bright everywhere the background should be, it is not always dark where the foreground should be.  Thus, the matting algorithm somehow needs to remove the foreground appearance from the green channel based on the red and blue channels.  As described by \citet{Smith:1996:Blue}, this can be done if the color space of the foreground object is limited to at most two dimensions, such as when it is known to be neutral in color, or flesh-toned, or lacking in the color present in the background.  Both film-based \citep{Vlahos:1964:Composite} and digital compositing techniques used in commercial products such as {\em Ultimatte} have assumed a limited foreground object color space, called the Vlahos Assumption.

\subsection{Recording a Separate Alpha Channel}

Another technique which has been explored is to photograph the alpha channel as a fourth color channel, simultaneous to the RGB color channels, using a reserved part of the spectrum.  {\em Infrared matting} places the actor in front of a visibly black screen reflecting infrared (IR) light, and uses beamsplitter to direct the IR toward a separate strip of film \citep{Pickley:1946:CP} or digital video camera, e.g., \citep{Debevec:2002:Lighting}.  The infrared image sees the actor silhouetted against a bright background, forming a holdout matte, or the inverse of the alpha channel.  Infrared matting is challenged by the fact that infrared light tends to focus differently through optics than visible light \citep{Vidor:1960:Infrared}.  The sodium vapor matting process \citep{Vlahos:1958:CPU} moves the matting channel into the visible spectrum by filming the actor in front of a field of monochromatic yellow-orange light from a low-pressure sodium vapor lamp.  A beamsplitter and bandpass filter allows the yellow background to be filmed on one strip of film, forming a holdout matte, and the actor's full-color appearance to be recorded through a notch reject filter on a second strip of film, blocking the sodium vapor light to show the actor normally lit on a black background.  Our work takes some inspiration from sodium vapor matting, but we eliminate the need for a specialized camera by using the green channel (rather than infrared or yellow) to record the matte, and we use a deep learning colorization algorithm to restore the missing green channel of the actor.  Numerous other matting techniques employing specialized optical properties have been proposed and used, including polarization \citep{Ben-Ezra:2000:Segmentation} and retroreflectivity \citep{Jenkins:1952:AFP}.

\subsection{Natural Image Matting}

Natural image matting is the process of separating a foreground element from a complex background, where the background color is variable and sometimes unknown across the image.  Numerous techniques from Bayesian Matting \citep{Chuang:2001:Bayesian} onwards \citep{Levin:2007:Closed,Levin:2008:Spectral,Sun:2004:Poisson,Wang:2007:Simultaneous}
have proposed automated and semi-automated techniques for natural image matting.  Typically, these algorithms begin with a manually drawn segmentation into known foreground ($\alpha = 1$), known background ($\alpha = 0$), and unknown transition (0 < $\alpha < 1$) regions, and the algorithm estimates $\alpha$ and the foreground/background colors $F$ and $B$ at each pixel in the transition region.  These algorithms have made their way into software tools available for production use, but still typically require manual input to achieve good results.

\subsection{Matting with Deep Learning}

Recently, deep learning techniques have been applied to the natural image matting problem with significant success, as in \citep{Xu:2017:Deep,Lin:2021:Real,Cai:2019:Disentangled,Chen:2018:TOM,Hou:2019:Context,Li:2020:Natural,Lutz:2018:AlphaGAN,Sengupta:2020:Background,Zhang:2021:Deep,Forte:2020:FBAlpha}.  However, an impediment to achieving production-quality alpha channels with these techniques has been the lack of training datasets with accurate ground truth alpha channels.  Many such datasets used in research use roughly estimated alpha channels from actors in front of a green screen, with inaccurate alpha variation in transition regions and foreground elements which have not been well-separated from the background color.  Our work is not in the category of natural image matting algorithms, as we film our actors in front of a green screen, but a major motivation of our technique is to generate accurate ground truth alpha channel data for training natural image matting algorithms.

\subsection{Image Colorization}

Image colorization has a similarly rich history to image matting and is typically framed as converting a single-channel grayscale image to RGB color, inferring three color channels from just one.  While early colorization techniques required manual digital painting, recent techniques surveyed by \citet{Anwar:2020:Image} and \citet{Vzeger:2021:Grayscale} leverage machine learning to make better guesses as to the colors in the original imagery.  Techniques from before the era of deep learning use statistical analysis or user-supplied color examples to colorize an image \citep{Reinhard:2001:Color,Welsh:2002:Transferring,Levin:2004:Colorization,Liu:2008:Intrinsic,Chia:2011:Semantic}.  

Deep learning--based colorization techniques \citep{Cheng:2015:Deep,Iizuka:2016:Let,Zhang:2016:Colorful,Isola:2017:Image,Deshpande:2017:Learning,Zhang:2017:Real,He:2018:Deep,Yoo:2019:Coloring,Su:2020:Instance,Kumar:2021:Colorization,Saharia:2022:Palette,Huang:2022:Unicolor} have leveraged vast image collections as training examples for supervised learning, as any color RGB image is easily converted to monochrome to form a training pair.  Nonetheless, \citet{Su:2020:Instance} note that ``image colorization is inherently an ill-posed problem with multi-modal uncertainty.'' If a monochrome image shows a person wearing a grey shirt, it's rarely clear what color the shirt should actually be.
\citet{Limmer:2016:Infrared} notably infer visible color RGB images from infrared imagery, using a spectral channel disjoint from the visible spectrum.

Deep colorization techniques have also been applied to video, emphasizing the need to achieve temporal stability \citep{Vondrick:2018:Tracking,Zhang:2019:Deep,Lei:2019:Fully}.

Informed by these works, we employ a colorization technique to restore the green channel to imagery containing only the red and blue channels of the scene.  As this requires inferring one channel from two, we have a simpler problem than monochrome-to-color inference, and we find that an example-based approach works well.

\section{Studio Lighting Setup}

\begin{figure}
    \centering
    \includegraphics[width=\columnwidth]{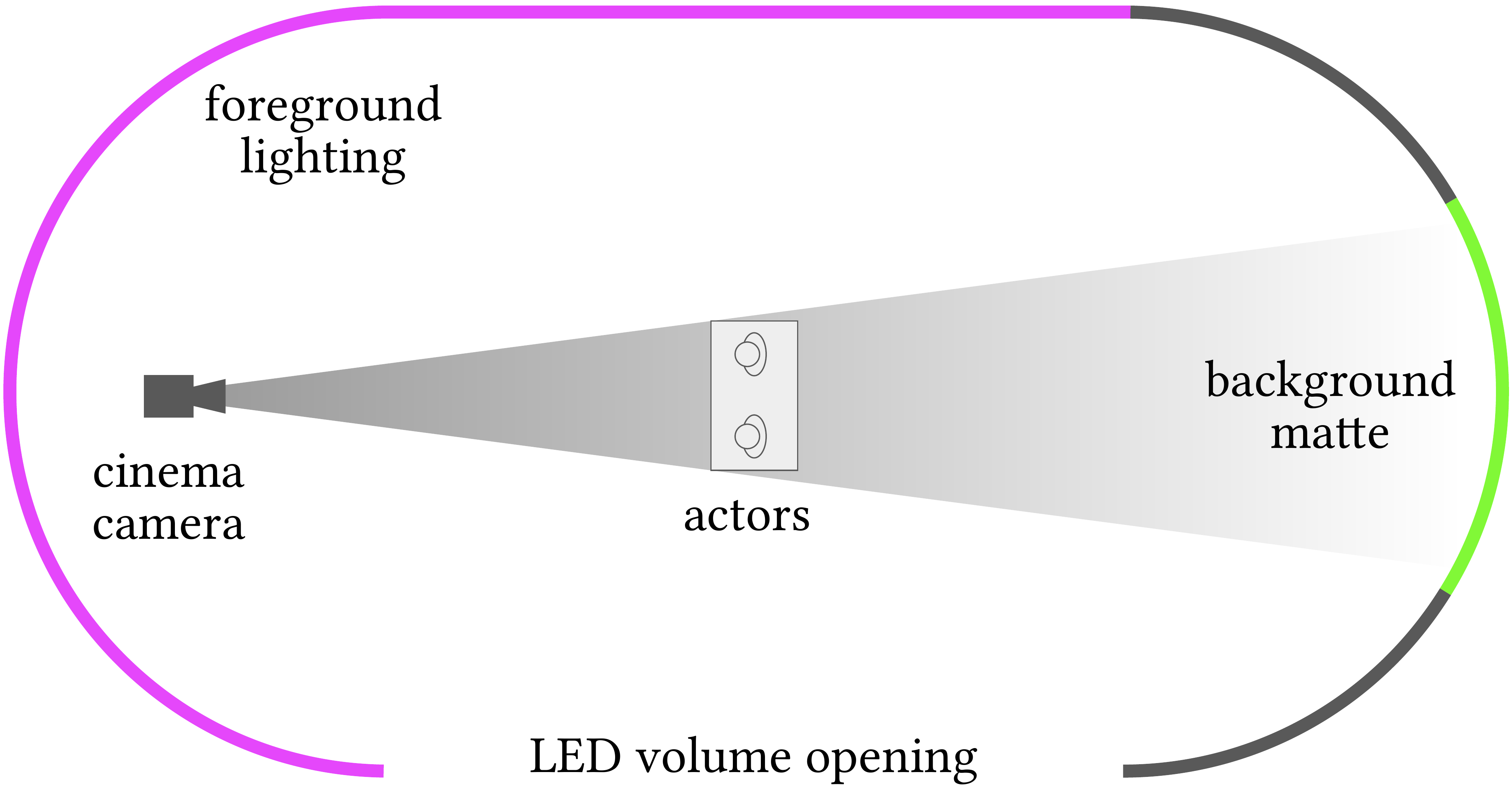} \\
    \vspace{1em}
    \includegraphics[width=\columnwidth]{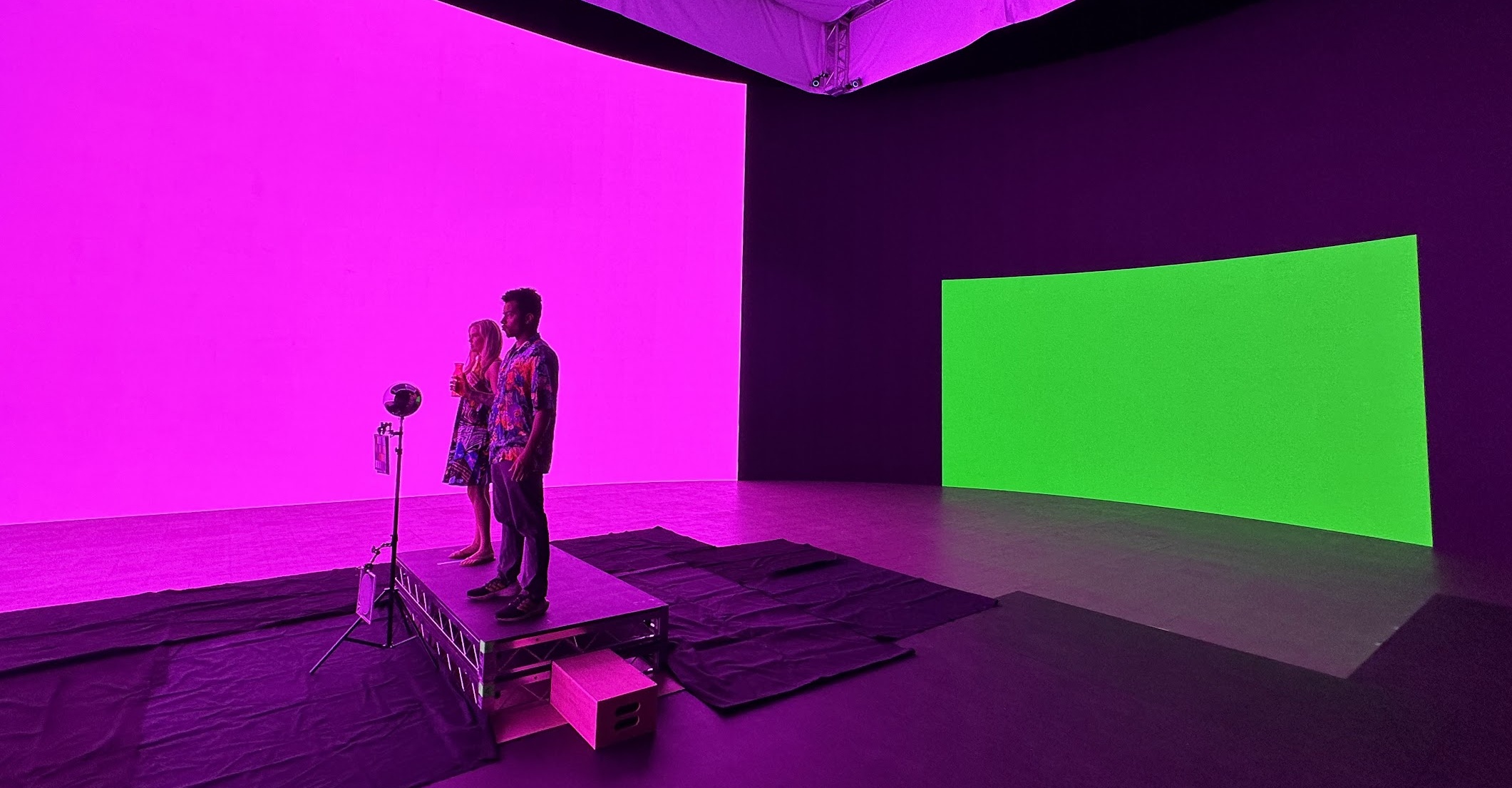}
    \vspace{-1em}
    \caption{A diagram of our principal LED volume filming setup with magenta foreground lighting and a green screen background within the camera frustum (top), and a photo of the setup (bottom).}
    \label{fig:setup}
\end{figure}

We filmed our actors in an LED volume \citep{Hamon:2014:Gravity,Bluff:2020:ILM} 18 meters wide and 9 meters deep, as shown in \Cref{fig:setup}.  The volume surrounds the actors 270 degrees around with ROE Black Pearl 2v2 LED panels, each being 50cm on a side with $176 \times 176$ pixels for a 2.8mm pixel pitch.  The panels consist of red, green, and blue LEDs.

The actors stood on a platform in the middle of the stage facing toward one curved side wall, with the other curved side wall behind them.  We used the walls in front and to the side of the actor for lighting and an area of the wall just behind the actors for the background, cropped tightly around the camera frustum to minimize spill light.  In the canonical configuration, we drove the lighting with a magenta color consisting of only the red and blue LEDs and the background with a green color consisting of only green LEDs.  To validate the technique, various foreground and background images were also placed on the lighting walls and inside the background camera frustum.

We filmed our subjects with a RED Komodo digital cinema camera\footnote{\url{www.red.com/komodo}} commonly used for digital filmmaking.  We filmed on a Canon 50mm EF lens set to an f/2.8 aperture to avoid the appearance of moir\'e in the background.  For the non-time-multiplexed recordings, we set the frame rate to 24fps and the shutter angle to 180 degrees to yield the typical frame rate and amount of motion blur commonly seen in movies.  For the time-multiplexed recordings, we shot at 48fps with a significantly narrower 105 degree shutter angle, with successive frames recording alternating lighting conditions, as described in \Cref{sec:time-multiplexing}.

Our subjects were chosen to have differing skin tone, hair color, and hair length and were costumed colorfully to challenge the colorization algorithm.  Each subject was handed a glass bottle, one red, one green, to test the algorithm's ability to record semitransparent alpha values.  They were directed to perform actions which showed the costumes from different angles and produced regions with significant motion blur.  A mirrored sphere and color chart was placed in the scene to document the lighting.  The reflection in the mirrored sphere gives an indication of the foreground lighting.  Lighting the subjects from the front and the left side of the image, but not the right, produced different matting challenges on their left and right sides.

While any cinema camera delivering linear pixel values should work with the basic magenta-green technique, we used the Komodo, since it is a global shutter camera, capable of synchronizing to the changing lighting conditions of the time-multiplexed matting techniques described in \Cref{sec:time-multiplexing}.

Our matting technique can also be practiced without an LED volume.  For example, our initial experiments were performed using inexpensive RGB LED light wands, lighting the actor with two wands set to a magenta color and lighting a white background with two wands set to a green color.

\section{Basic Method}

Our method honors Alvy Ray Smith's assertion that the ``transparency of an image is as fundamental as its color'' \citep{Smith:1995:Alpha} and uses one of the camera's three color channels (usually green) to measure the alpha channel.  We then use example-based image colorization to restore the green channel of the foreground element.

We can explain this process in terms of the matting equations \citep{Porter:1984:Compositing}.  We refer to a pixel's background color using the RGB triple $[B_R, B_G, B_B]$, its foreground subject as $[F_R, F_G, F_B]$, and its composited appearance as $[C_R, C_G, C_B]$. Assuming a single alpha transparency $\alpha$ for all color channels, the matting equations are simply

\begin{equation}
\begin{split}
C_R &= \alpha F_R + (1-\alpha)B_R \\
C_G &= \alpha F_G + (1-\alpha)B_G \\
C_B &= \alpha F_B + (1-\alpha)B_B.
\end{split}
\label{Eqn:MattingEqns}
\end{equation}

These equations include seven total unknowns for a given photograph: $B_R$, $B_G$, $B_B$, $F_R$, $F_G$, $F_B$, $\alpha$, as the photograph's pixel values comprise $C_R$, $C_G$, $C_B$. In the case that the background color $B_R$, $B_G$, $B_B$ can be measured (e.g., by photographing a clean plate without the foreground subject) there are four unknowns: $F_R$, $F_G$, $F_B$, $\alpha$. While many alpha matting algorithms focus on inferring $\alpha$, it should be clear from these equations that to form a successful image composite over a new background that indeed the foreground color $F_R$, $F_G$, $F_B$ must also be recovered, leaving the problem ill-posed without additional constraints.

\citet{Smith:1996:Blue} noted that if the subject reflects no blue light, then $F_B = 0$, and the blue channel of the subject in front of a blue screen gives a direct measurement of $1-\alpha$, which allows $F_R$, $F_G$, and $\alpha$ to be determined easily.  We leverage color-controllable RGB LED lighting to a similar end: we turn off the green LEDs lighting an arbitrary subject to force $F_G = 0$ and illuminate them from behind with a field of green light.  In this way,

\begin{equation}
\begin{split}
C_R &= \alpha F_R + (1-\alpha)B_R \\
C_G &= (1-\alpha)B_G \\
C_B &= \alpha F_B + (1-\alpha)B_B.
\label{Eqn:MattingEqnsNoGreenForeground}
\end{split}
\end{equation}

Rearranging to solve for the three remaining unknowns yields

\begin{equation}
\begin{split}
\alpha &= \frac{B_G - C_G}{B_G} \\
F_R &= \frac{C_R - (1-\alpha)B_R}{\alpha} \\
F_B &= \frac{C_B - (1-\alpha)B_B}{\alpha}.
\label{Eqn:MattingEqnsNoGreenForegroundRearranged}
\end{split}
\end{equation}

Note that these equations are solvable only if $B_G > 0$, as otherwise the first equation is undefined. Furthermore, if $\alpha$ is zero, the foreground colors $F_R$ and $F_B$ are undefined. The intuition behind these equations is that the green channel is guaranteed \textit{only} to be nonzero in the background, and so it is now essentially just a silhouette image of the subject, with pixel values of zero everywhere in the foreground. This is the inverse of the desired alpha channel, up to a scale factor. Given this more intuitive interpretation, it is clear why the main additional constraint is that the background must contain green; otherwise no silhouette image remains.

In this designed scenario, the foreground is guaranteed to have $F_G = 0$. However, we could have also suggested that $F_R = 0$ or $F_B = 0$, implying no red or blue pixels in the foreground, respectively (see \Cref{sec:yellowblue}). 



\subsection{Color Calibration}
\label{sec:crosstalk}

Digital cinema cameras sense color by placing a color filter array---often a Bayer pattern---over a set of photosites sensitive to the entirety of the visible spectrum.  The color filters, by design, usually have a significant degree of overlap in spectral transmission.  As a result, a given wavelength of red light may register on both the red and green color channels of an image, a phenomenon known as {\em crosstalk}.  




\begin{figure}
    \centering
    \includegraphics[width=\linewidth]{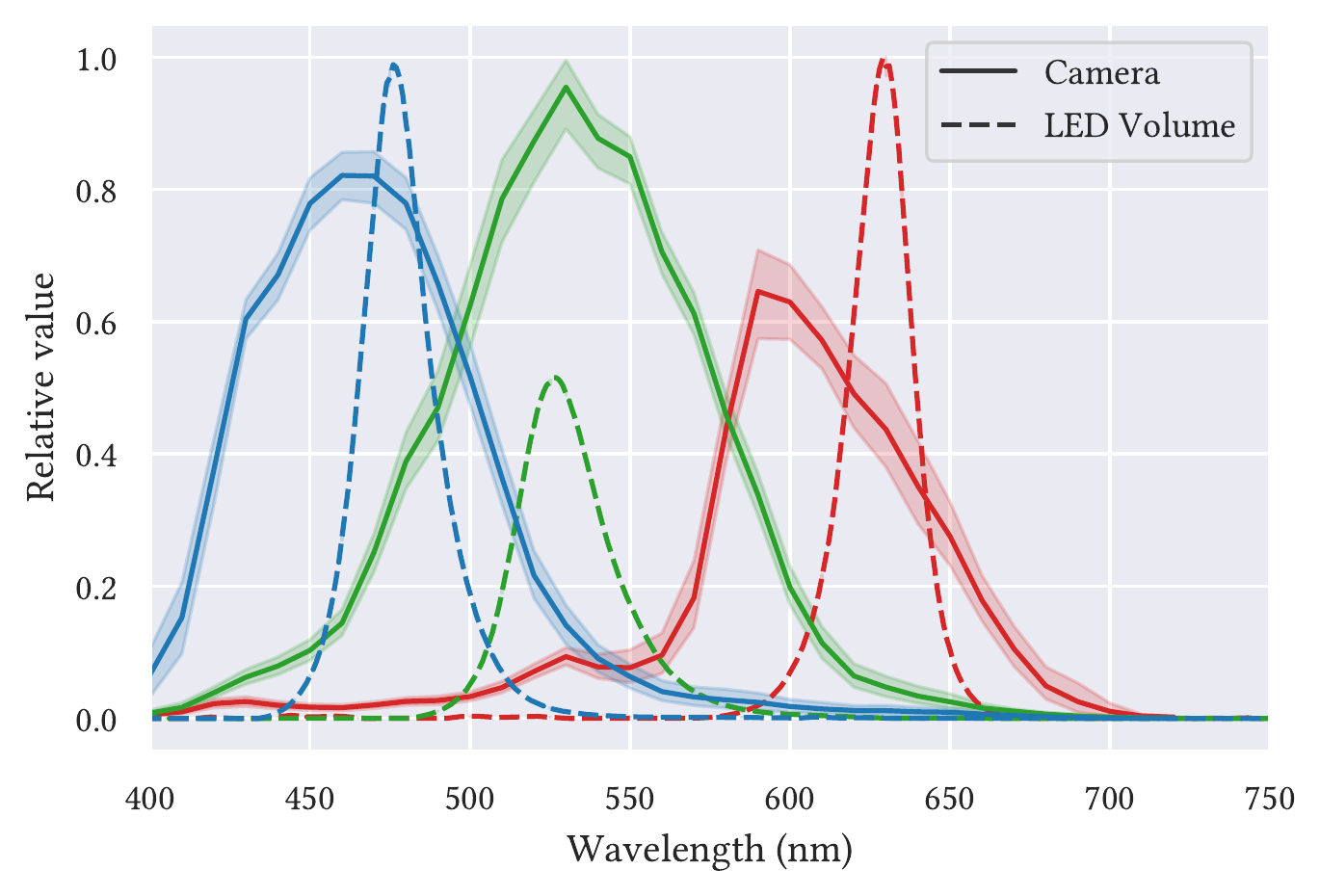}\vspace{-1em}
    \caption{The spectral sensitivity curves of a variety of measured camera sensors \citep{jiang2013specular} overlaid with the spectral output of the red, green, and blue LEDs of the LED panels in an LED volume, showing crosstalk.}
    \label{fig:spectral}
\end{figure}

Since our method as outlined in \eqref{Eqn:MattingEqnsNoGreenForegroundRearranged} requires that the green pixel values of the foreground content are all zero and, ideally, that the red and blue channels show the subject against black, we need to remove this color crosstalk.  This can be done effectively with a $3 \times 3$ color transformation matrix $\mathbf{M}$.  To determine $\mathbf{M}$, we first record the appearance of each of the LED spectra to the cinema camera by placing a color chart in the scene and illuminating it consecutively by red, green, and blue light.  We calculate the average RGB color of the chart's white square under each lighting condition and places these RGB values as column vectors into a measurement matrix $\mathbf{W}$.  The matrix records how much of each LED color affects each color channel. Since $\mathbf{W}$ transforms the individual LED colors to camera observations, $\mathbf{M=W^{-1}}$ transforms camera observations back to the individual LED colors, removing the crosstalk.  We thus apply this color calibration matrix $\mathbf{M}$ to all imagery prior to applying the solution of \eqref{Eqn:MattingEqnsNoGreenForegroundRearranged}. 

This calibration process allows us to pretend that we captured our imagery using a camera with ``sharp'' spectral sensitivities without color channel crosstalk.  This solution is only guaranteed for materials of spectrally neutral reflectance, such as the white square of the color chart, since having a light spectrum reflect off of a material reflectance spectrum transforms the spectral content of the observed light, and a somewhat different crosstalk elimination matrix could be required for different materials.  However, since the spectral output of the LEDs in a virtual production stage is narrow (\Cref{fig:spectral}), the crosstalk ratios tend to remain similar for the majority of the spectral content of each LED.  In practice, we are able to eliminate the majority of crosstalk even for strongly hued materials with a single $\mathbf{W}$ applied to the whole image.

\subsection{Bounce Light Subtraction}
\label{sec:bouncelight}

\begin{figure}
     \centering
     \begin{subfigure}[t]{0.48\columnwidth}
         \centering
         \includegraphics[width=\textwidth]{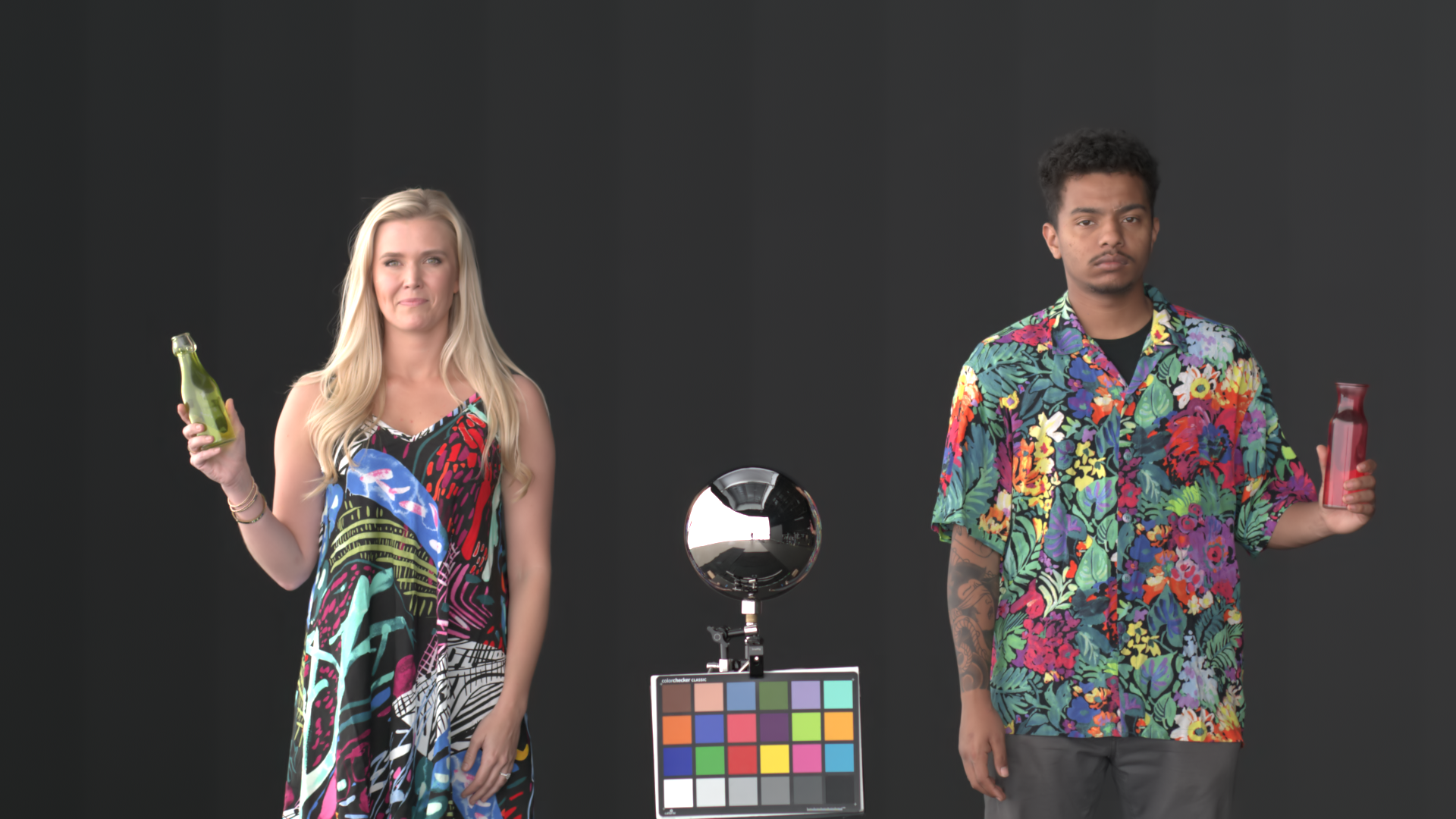}
         \caption{}
         \label{fig:bounce-actors}
     \end{subfigure}
     \hfill
     \begin{subfigure}[t]{0.48\columnwidth}
         \centering
         \includegraphics[width=\textwidth]{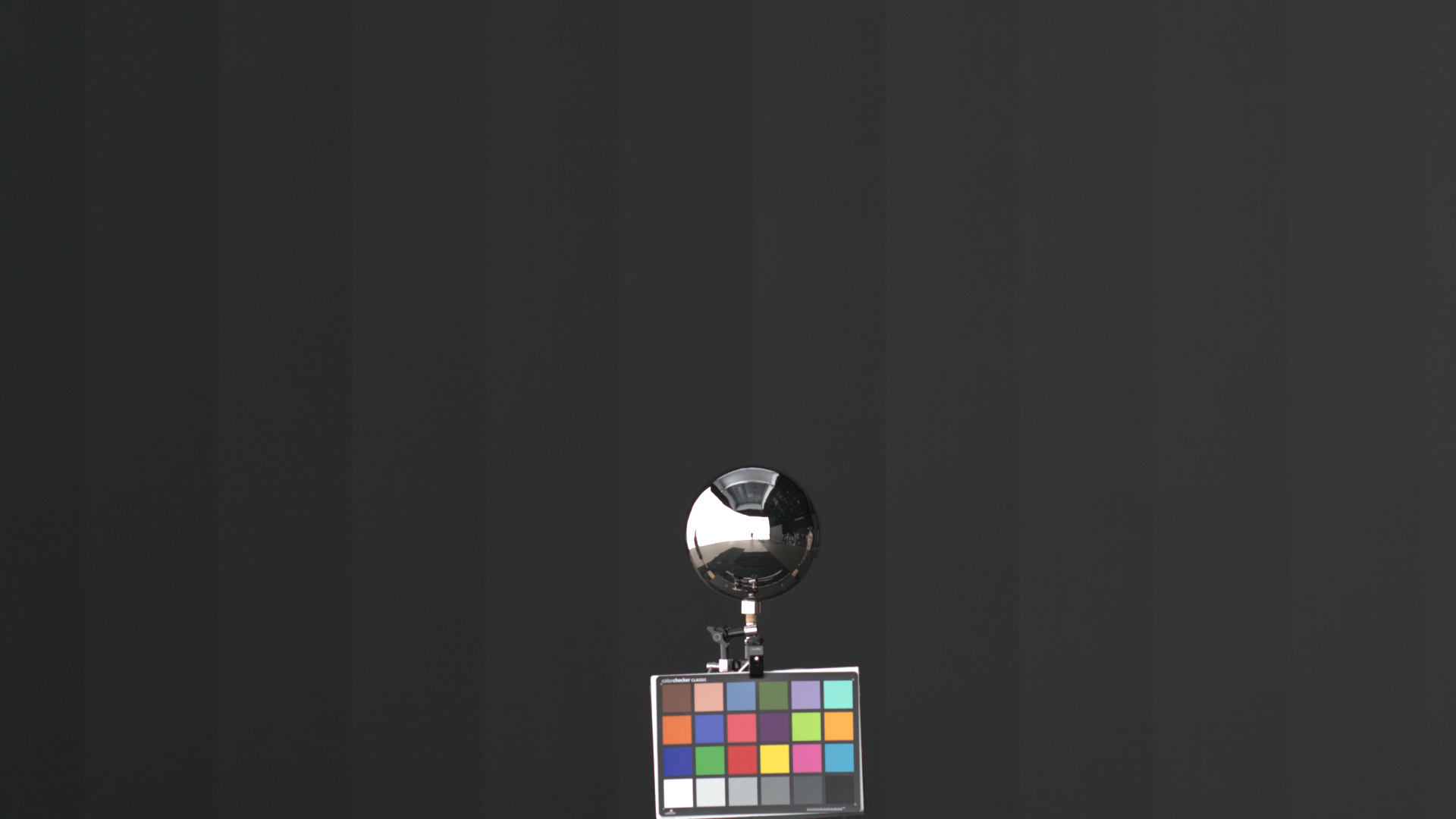}
         \caption{}
         \label{fig:bounce-clean}
     \end{subfigure} \\
     \begin{subfigure}[t]{0.48\columnwidth}
         \centering
         \includegraphics[width=\textwidth]{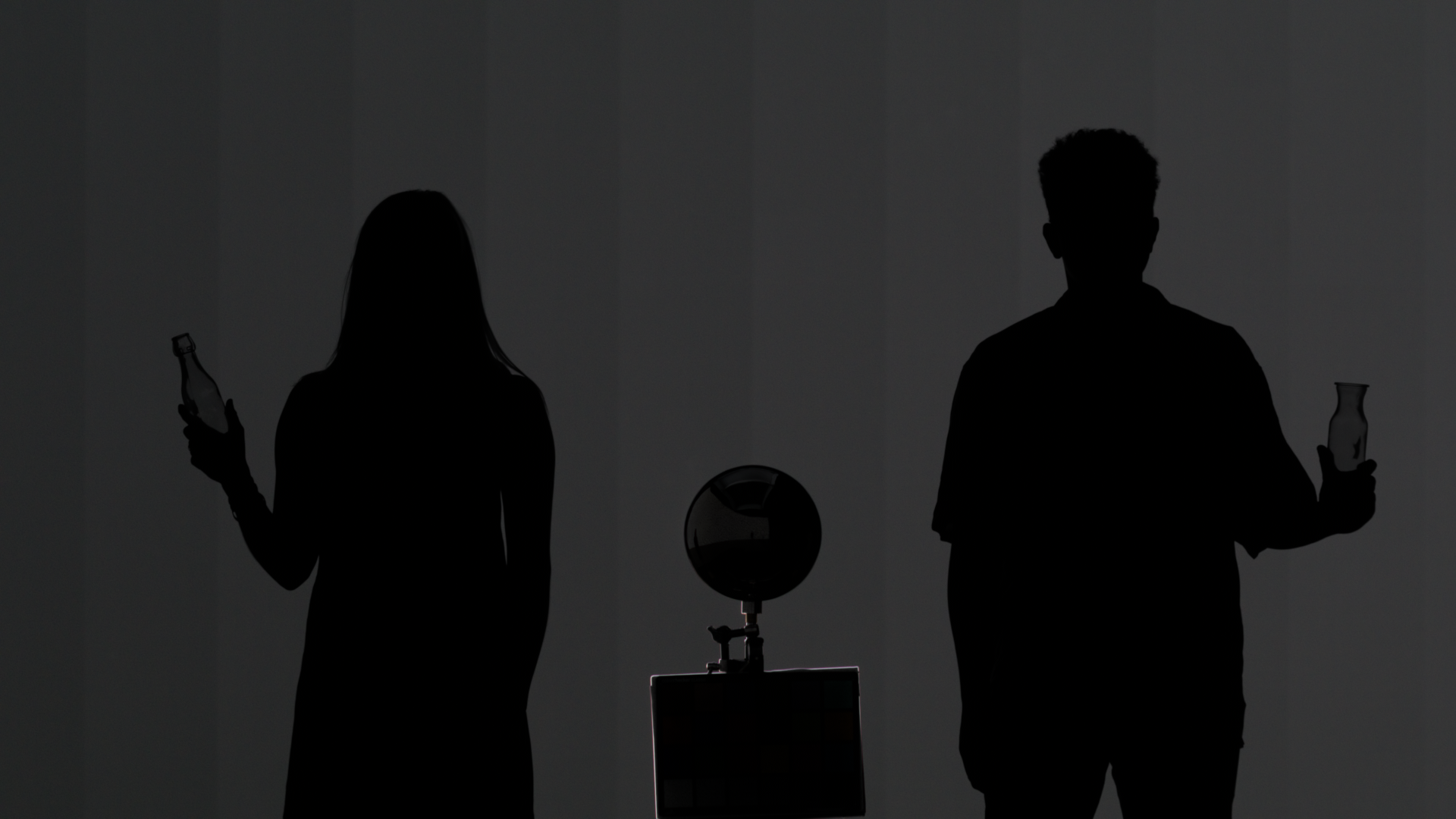}
         \caption{}
         \label{fig:bounce-holdout}
     \end{subfigure}
     \hfill 
     \begin{subfigure}[t]{0.48\columnwidth}
         \centering
         \includegraphics[width=\textwidth]{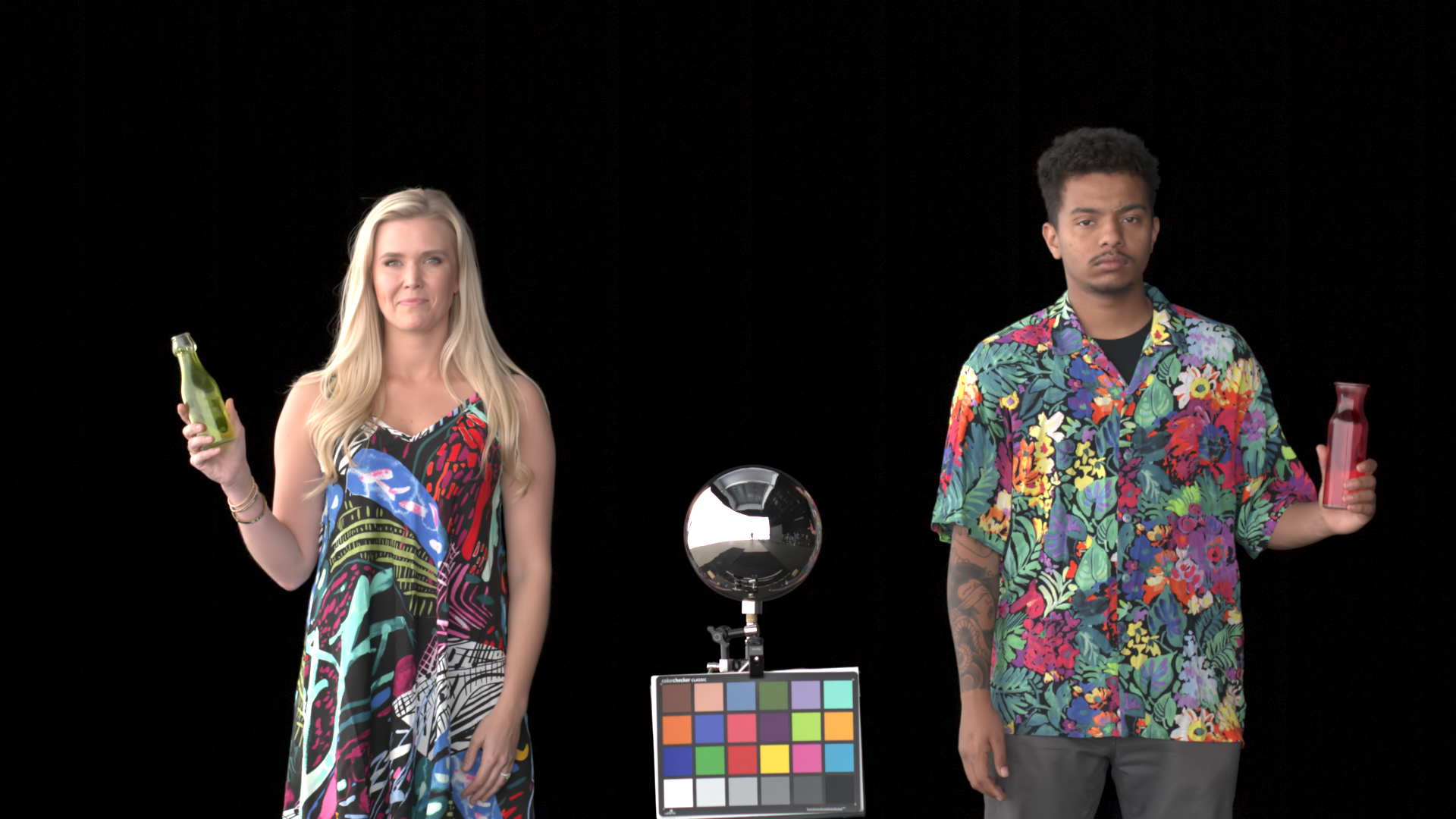}
         \caption{}
         \label{fig:bounce-corrected}
     \end{subfigure}
     \vspace{-0.5em}
        \caption{Correcting for bounce light. Actors are captured in front of an unlit background exhibiting bounce light (a). We also capture a clean plate showing the bounce light on its own (b).  We subtract the clean plate multiplied by the holdout matte (c) from (a) to remove bounce light on just the background, not the actors (d).}
        \label{fig:bounce}
\end{figure}

A final pre-processing step is necessary is to correct for the presence of bounce light within the LED volume.  Because LED panels are not perfectly black and reflect some of the light falling upon them, the background panels behind the actors will typically include some bounced light from the foreground.  This means that the foreground element will not be seen against a perfect field of black, as is required for the element to be self-matting with a premultiplied alpha but against a field of dim reflected foreground light, a seen in \Cref{fig:bounce-actors}.  We can measure this bounced light by turning off the background LED panels while they are illuminated by foreground lighting as in \Cref{fig:bounce-clean}.  After color correction, we can then subtract this bounced light from just the background around the actors by first multiplying it with the holdout matte as in \Cref{fig:bounce-holdout}, yielding the result in \Cref{fig:bounce-corrected}.

\subsection{Colorizing the Missing Green Channel}
\label{sec:colorize}

While the \textit{Magenta Green Screen} process records an accurate alpha matte, the resulting foreground elements have the serious deficiency of missing their green channel.  To address this, we design an image colorization technique to restore the green channel based on the observed red and blue channels.

\subsubsection{Naive Colorization}

Real-time colorization can be performed in a simple but naive way by setting the green channel to be a linear combination of the red and blue channels: $g = \rho r + (1-\rho) b$.  The value $\rho$ can be $0.5$ to average the channels or can be chosen to be a value which optimizes the appearance of skin tones, closer to $\rho=0$.  Such naively-colorized images have a limited and innacurate range of colors, but they can get skin tones, neutral tones, blue skies, and dusty plains all to look approximately correct, which explains why the two-strip Technicolor process could be effectively used for Westerns.  However, greens, magentas, and many other colors will be represented inaccurately.  \Cref{fig:teaser-naive} shows a naively-colorized image from the magenta green matting process.

\subsubsection{Colorization with Deep Learning}

To recover the green channel more accurately, we can train a deep neural network to infer the green channel from the red and blue channels based on full-color training examples.  For this we follow previous works which perform full colorization of grayscale images, e.g., \citep{Zhang:2016:Colorful}, emboldened by the knowledge that our problem is a significantly easier one, restoring one channel of information from two rather than two from one.

\begin{figure}
    \centering
    \includegraphics[width=0.4\textwidth]{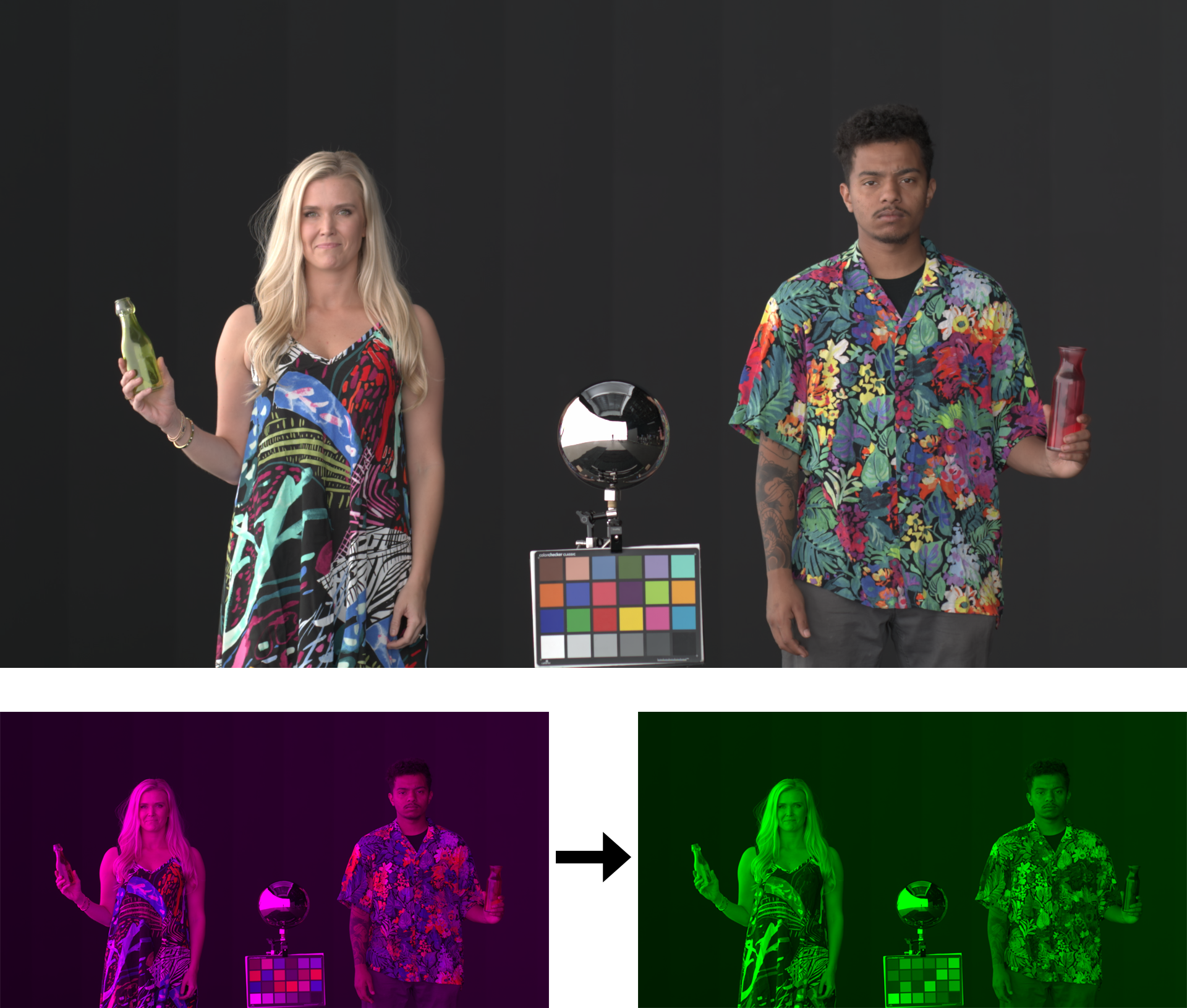}
    \vspace{-0.5em}
    \caption{An example frame of training data used for our colorization model. Given frames of a subject captured with white lighting in front of a black background, we train a model to predict the green channel from the red and blue.}
    \label{fig:training-data}
\end{figure}

In this work, we record scene-specific training data in the form of an alternate ``rehearsal'' take of the scene shot under white RGB lighting seen on a black background, as seen in \Cref{fig:training-data}.  Each frame from this sequence yields a colorization training pair showing the proper green channel for given red and blue channels.  

During training, we take random $512 \times 512$ crops of the full $1920\times1080$ high definition frames and perform data augmentation by randomly perturbing the image luminance and color balance. Since our network is fully-convolutional, we can apply to it to the full-resolution image at test-time despite training on patches. We found that tone mapping the images with a gamma value of 2.2 improved the results compared to training on linear data, since doing the latter leads to poor optimization in darker areas of the images.

We use a standard image-to-image translation U-Net architecture with skip connections \citep{Ronneberger:2015:UNet}. We first 
pass the input through two $3\times3$ convolutional layers followed by five downsampling blocks, starting with 32 channels and doubling the number at each layer, five upsampling blocks with corresponding numbers of channels, two additional $3\times3$ convolutions, and a final $1\times1$ convolution and $tanh$ non-linearity to constrain the output pixel values to a reasonable range. We use a Leaky ReLU non-linearity \citep{xu2020reluplex} and Batch Normalization \citep{Ioffe:2015:Batch} after each convolutional layer except for the first two. Each downsampling and upsampling block contains two $3\times3$ convolutions and uses blur-pooling \citep{zhang2019shiftinvar}.

We train our network for \num{100000} iterations using Adam \citep{kingma2014adam} as the optimizer and a learning rate of 0.0001 with batch size 16. This takes approximately 2.5 hours on four NVIDIA A10G GPUs.  Color inference is much faster, taking less than one second per frame.

\subsubsection{Time-Multiplexing}
\label{sec:time-multiplexing}

A significant drawback of Magenta Green Screen is that the actors need to perform their scene under the unnatural illumination condition of magenta light.  We can disguise the appearance of the magenta illumination by rapidly alternating it with green illumination, with the camera synchronized to the illumination changes so that it records only the magenta-green conditions.  This is similar to the effect demonstrated in \citep{McDowall:2004:SI}, where black-and-white artwork was hidden within imagery projected by a high-speed video projector, with each artwork frame quickly followed by its black-and-white inverse.  Related time-multiplexing techniques have been shown for virtual production matting applications by \citet{Wenger:2005:PRR} and {\em GhostFrame\footnote{\url{https://www.ghostframe.com/}}}.

\begin{figure}
    \centering
    \includegraphics[width=0.35\textwidth]{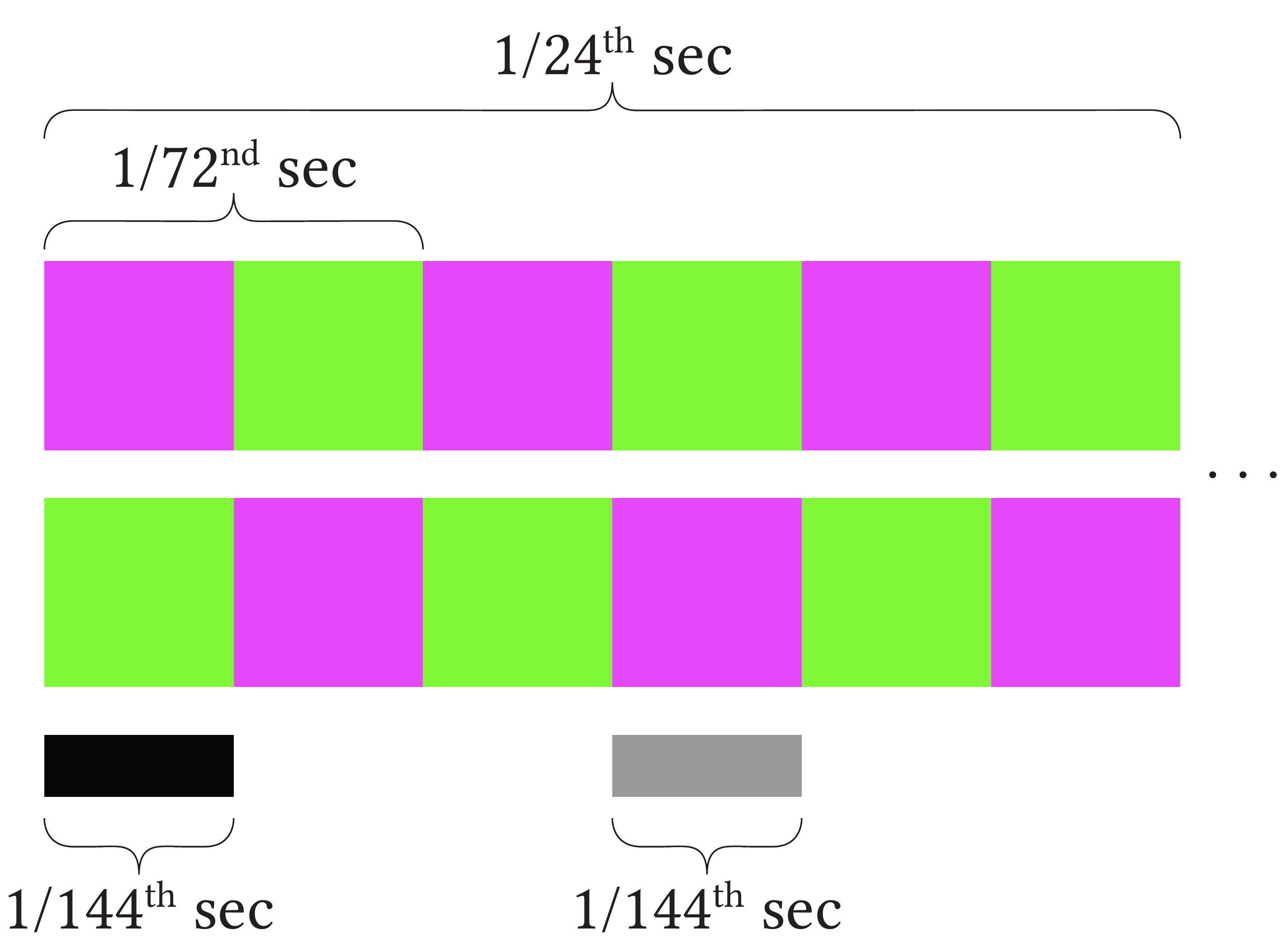} \\
    \vspace{0.1in} 
    \includegraphics[width=0.35\textwidth]{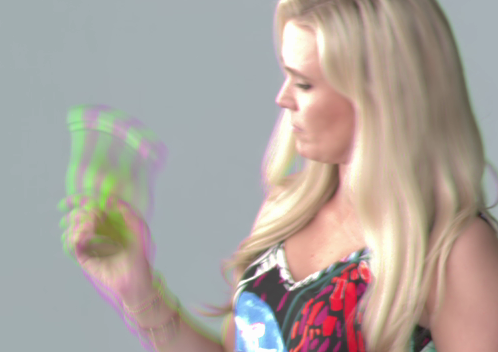}
    \vspace{-0.5em}
    \caption{Time-multiplexed version of Magenta Green Screen, repeating the pattern and its inverse at 72Hz (top). While the lighting appears neutral in color and does not visibly pulsate, fast-moving objects show color fringing to the eye, as approximated in this 360 degree shutter exposure (botton).}
    \label{fig:mgrainbow}
\end{figure}

Unfortunately, alternating between lighting patterns at 24Hz is uncomfortably stroboscopic and could even be dangerous for a person with a sensitivity to flashing light.  According to \citet{Fisher2022VisuallySS}, for such photosensitive individuals, ``images with flashes brighter than 20 candelas/$m^2$ at 3-60 (particularly 15-20) Hz occupying at least 10 to 25 degrees of the visual field are a risk."  We address this by increasing the repeating rate of the two lighting conditions at 72HZ, so that the lighting changes from one color to the next every 144th of a second.  The lighting then appears nearly constant, with a remaining effect being that rapidly moving objects leave a trail of magenta/green outlines when seen against the screen, as in \Cref{fig:mgrainbow} (bottom).  We can then synchronize our cinema camera to record the first of every six lighting changes, requiring a shutter angle of at most 60 degrees, yielding Magenta Green Screen images at 24fps, as seen in \Cref{fig:mgrainbow} (top).  The magenta-green frames can be colorized as before from a separate pass lit by full-spectrum lighting. 

\begin{figure}
     \centering
     \begin{subfigure}[t]{0.48\columnwidth}
         \centering
         \includegraphics[width=\textwidth]{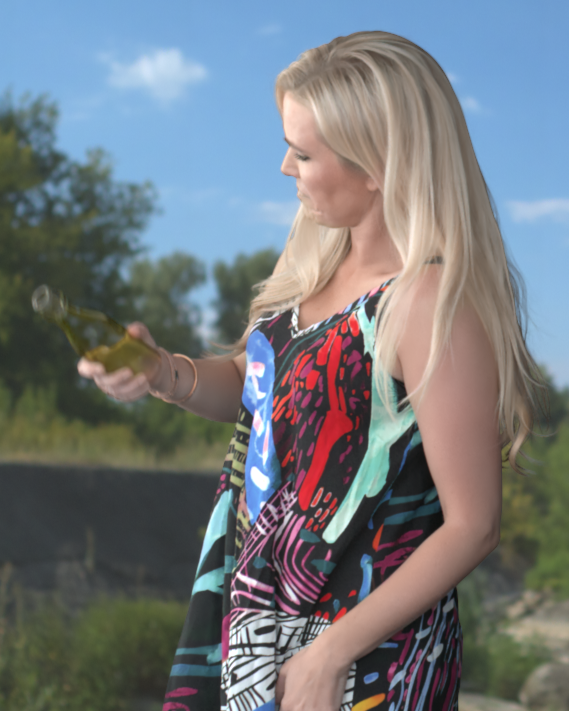}
         \caption{Original.}
     \end{subfigure}
     \hfill
     \begin{subfigure}[t]{0.48\columnwidth}
         \centering
         \includegraphics[width=\textwidth]{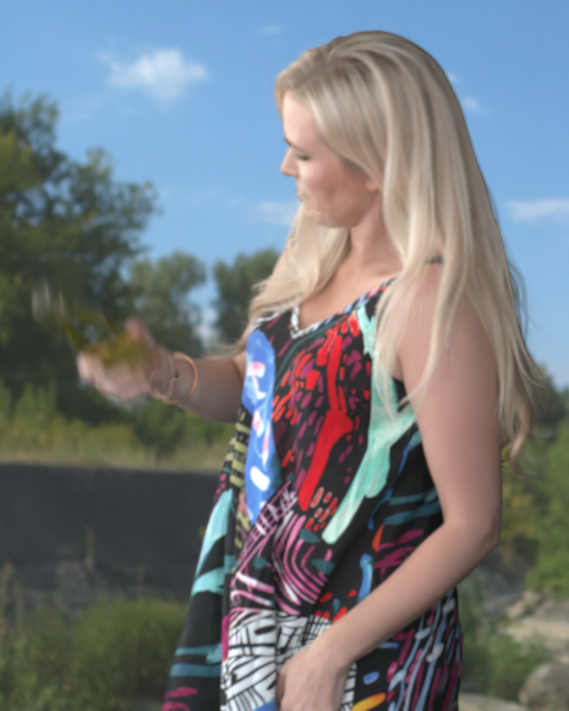}
         \caption{With simulated motion blur.}
     \end{subfigure}
     \vspace{-0.5em}
        \caption{We use optical flow to add simulated 180 degree motion blue to our Time-Multiplexed Magenta Green Screen composited footage.}
        \label{fig:motionblur}
\end{figure}

The remaining drawback of time-multiplexing in this manner is that the shorter shutter angle reduces the amount of motion blur, which is considered desirable for cinema. \citet{Wenger:2005:PRR} faced a similar problem in their time-multiplexed relighting work, with relit images formed as the linear combination of images taken with very short exposure times. Like this work, we can address the problem in our Time-Multiplexed Magenta Green Screen technique by computing optical flow and using the flow not only to temporally align adjacent frames, but also to add simulated 180 degree motion blur to the images. We show an example of this in \Cref{fig:motionblur} and in the accompanying video.

\subsection{Time-Multiplexed Magenta Green Screen}

\begin{figure}
     \centering
     \begin{subfigure}[t]{0.48\columnwidth}
         \centering
         \includegraphics[width=\textwidth]{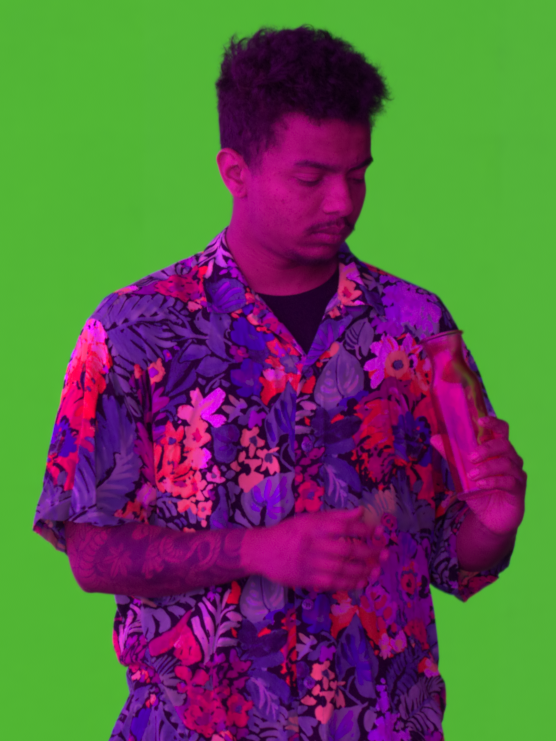}
         \caption{Magenta-green frame.}
         \label{fig:magenta-green-tm}
     \end{subfigure}
     \hfill
     \begin{subfigure}[t]{0.48\columnwidth}
         \centering
         \includegraphics[width=\textwidth]{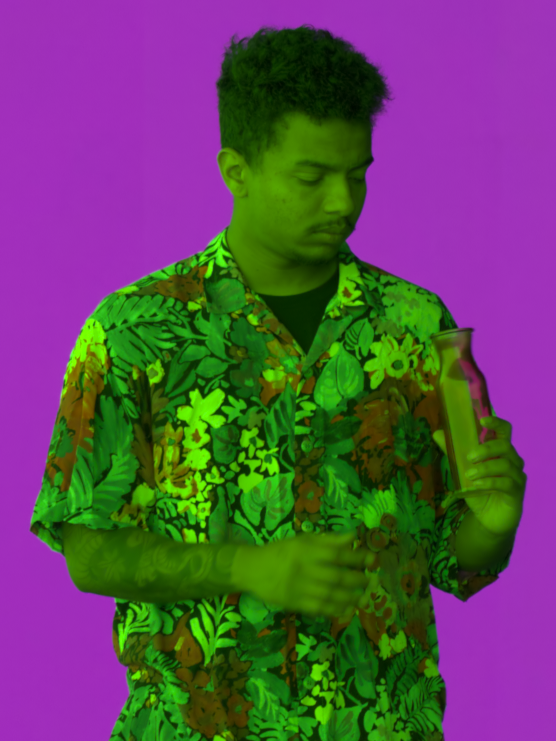}
         \caption{Green-magenta frame.}
         \label{fig:green-magenta-tm}
     \end{subfigure} \\
     \vspace{0.5em}
     \begin{subfigure}[t]{0.48\columnwidth}
         \centering
         \includegraphics[width=\textwidth]{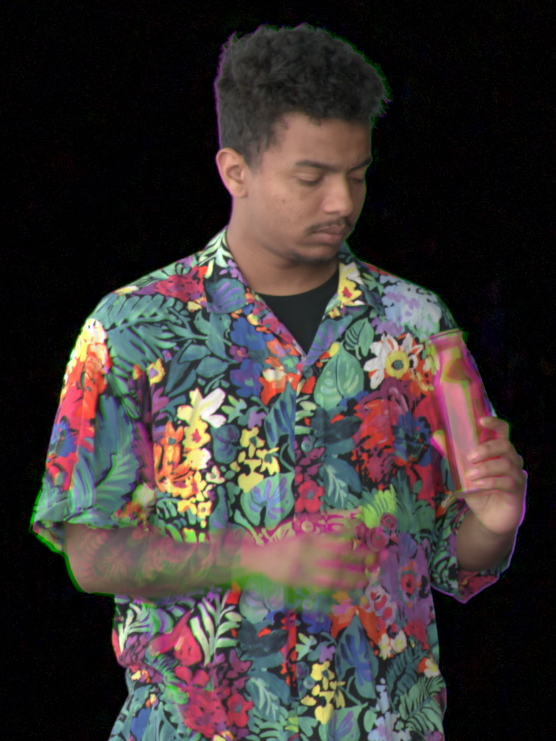}
         \caption{Foreground sans optical flow.}
         \label{fig:multiplexed-naive}
     \end{subfigure}
     \hfill 
     \begin{subfigure}[t]{0.48\columnwidth}
         \centering
         \includegraphics[width=\textwidth]{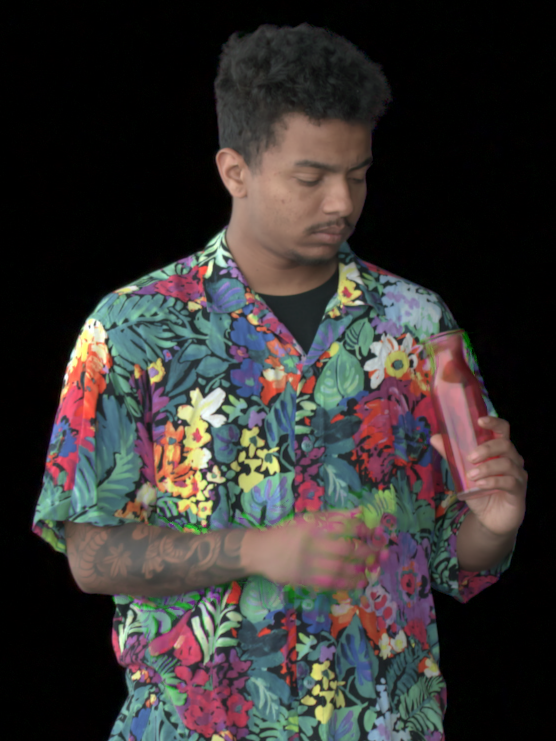}
         \caption{Foreground with optical flow.}
         \label{fig:multiplexed-flow}
     \end{subfigure}
     \vspace{-0.5em}
        \caption{In Time-Multiplexed Magenta Green Screen, a magenta-green frame (a) is quickly followed by a green-magenta frame (b) to neutralize its appearance.  We reconstruct the foreground naively (c) and using optical flow (d).}
        \label{fig:multiplexed}
\end{figure}

If we set the cinema camera to record at 48fps, it will record every third lighting condition in the time-multiplexed magenta-green sequence.  As seen in \Cref{fig:mgrainbow}, this yields alternating frames lit by magenta and green light, in front of a background of green then magenta light, as seen in \Cref{fig:magenta-green-tm,fig:green-magenta-tm}.  Notably, the second of each pair of frames contains the appearance of the actor under green light.  In the absence of motion, this green channel could complete the red and blue channels of the previous frame and eliminate the need for colorizing the foreground element.  However, in the presence of motion, there will be a frame misalignment, as seen in \Cref{fig:multiplexed-naive}.

We can attempt to align the frames by computing optical flow from one magenta-green frame to the next, and displacing the interposed green-magenta frame by half the estimated flow vectors, similar to the use of tracking frames to align intermediate lighting frames in \citep{Wenger:2005:PRR}.  We use the {\em Kronos} node in the Nuke studio compositing software for this purpose.  After applying motion compensation, we are able to recover the full RGB foreground element (\Cref{fig:multiplexed-flow}). However, when the subject moves quickly, the optical flow algorithm fails to track the motion, resulting in colorization errors.

\subsection{Classic Time-Multiplexed Matting }

\begin{figure}
     \centering
     \begin{subfigure}[t]{0.48\columnwidth}
         \centering
         \includegraphics[width=\textwidth]{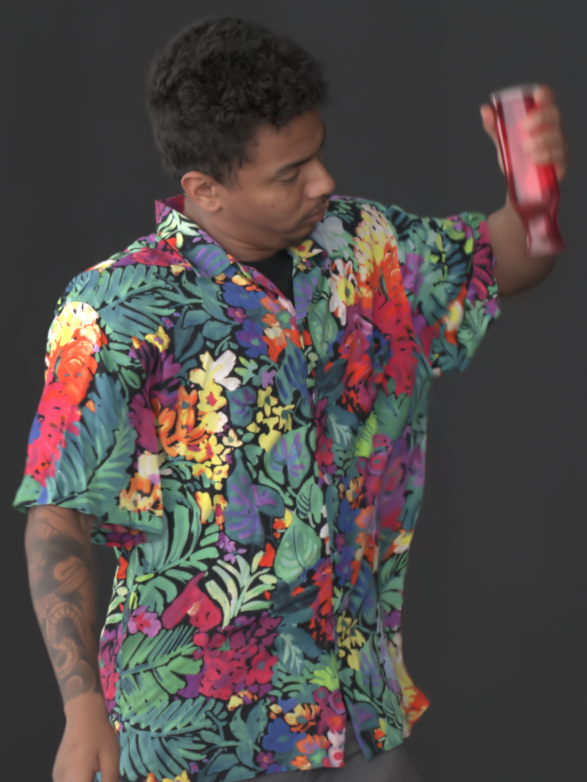}
         \caption{Foreground.}
         \label{fig:white-black-ctmm}
     \end{subfigure}
     \hfill
     \begin{subfigure}[t]{0.48\columnwidth}
         \centering
         \includegraphics[width=\textwidth]{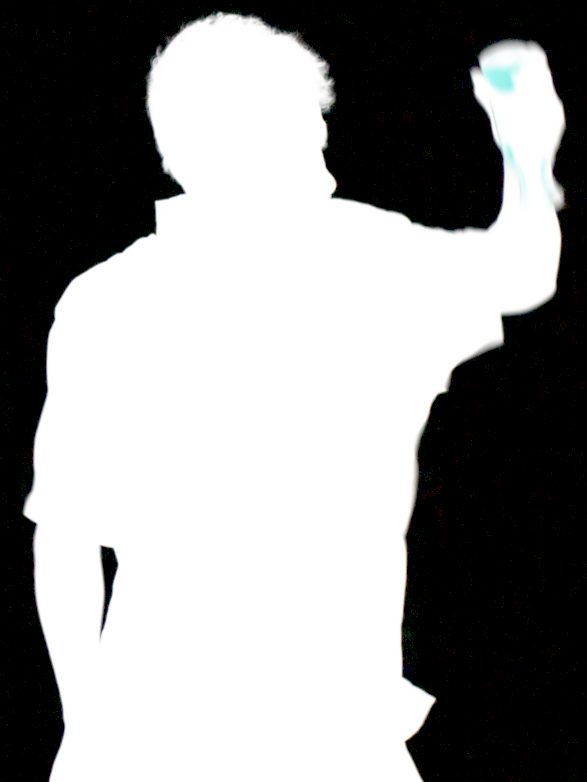}
         \caption{Matte.}
         \label{fig:black-white-ctmm}
     \end{subfigure} \\
     \vspace{0.5em}
     \begin{subfigure}[t]{0.48\columnwidth}
         \centering
         \includegraphics[width=\textwidth]{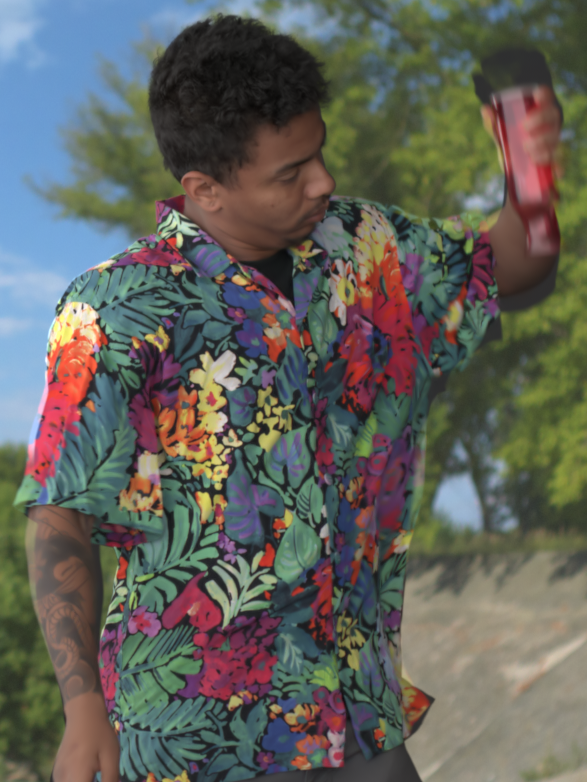}
         \caption{Composite sans optical flow.}
         \label{fig:ctmm-noflow}
     \end{subfigure}
     \hfill 
     \begin{subfigure}[t]{0.48\columnwidth}
         \centering
         \includegraphics[width=\textwidth]{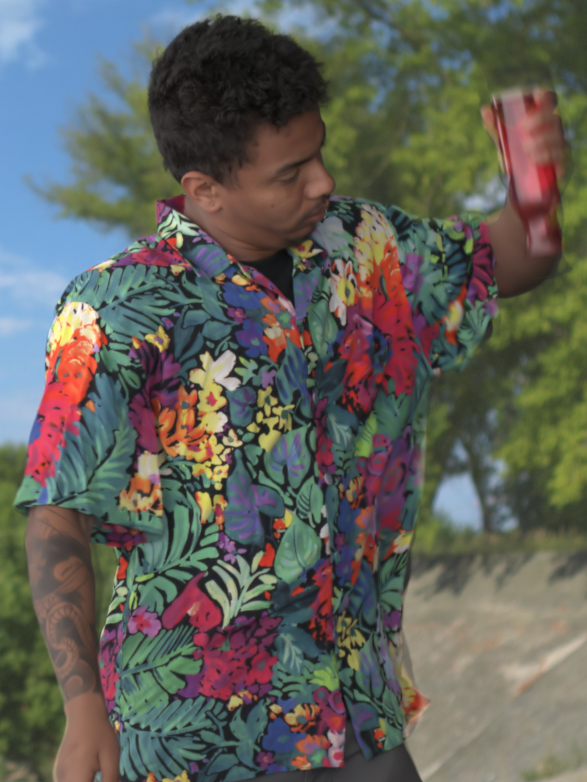}
         \caption{Composite with optical flow.}
         \label{fig:ctmm-flow}
     \end{subfigure}
     \vspace{-0.5em}
        \caption{We use consecutive frames from Classic Time-Multiplexed Matting (a, b) to reconstruct the foreground without optical flow (c) and with optical flow.}
        \label{fig:classicTMM}
\end{figure}

\citet{Wenger:2005:PRR} performed time-multiplexing matting by alternating frames lit by white light against a dark background with frames of the actor in silhouette against an illuminated background.  The technique required optical flow, and produced good results in standard definition video.  For comparison, we implemented this classical time-multiplexed matting approach by alternating between the actor lit by RGB white lighting against black and then unlit against an RGB white background, as in \Cref{fig:classicTMM}.  Applying optical flow between the illuminated frames to align the matte can produce a good composite and has the benefit of yielding a full-color matte frame, able to record color-dependent transparency.  However, as can be seen in the accompanying video, the optical flow can fail in the presence of fast subject motion, creating matte edges that are misaligned with the foreground element.  Our technique of shooting the matte in the same frame as the foreground element (which becomes colorized) does not require optical flow for alignment.

\subsection{Time-Multiplexed Triangulation Matting}

\begin{figure}
     \centering
     \begin{subfigure}[t]{0.495\columnwidth}
         \centering
         \includegraphics[width=\textwidth]{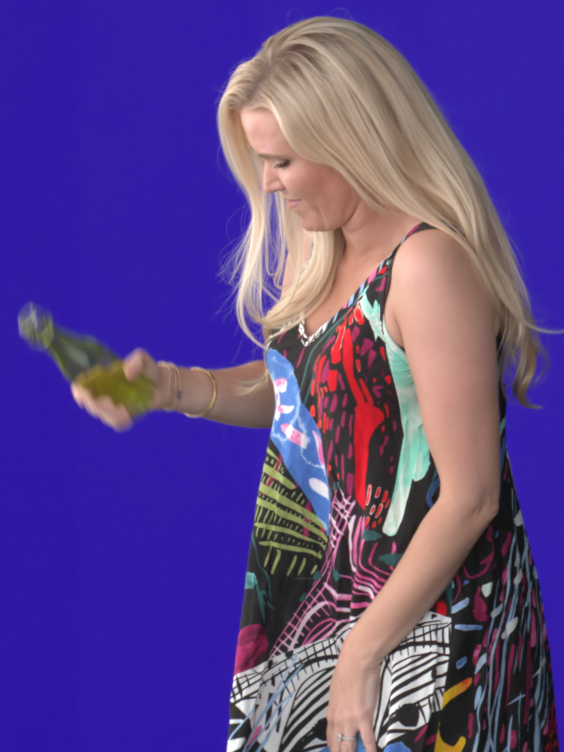}
         \vspace{-1.5em}
         \caption{Blue background.}
         \label{fig:blue-TMTM}
     \end{subfigure}
     \begin{subfigure}[t]{0.495\columnwidth}
         \centering
         \includegraphics[width=\textwidth]{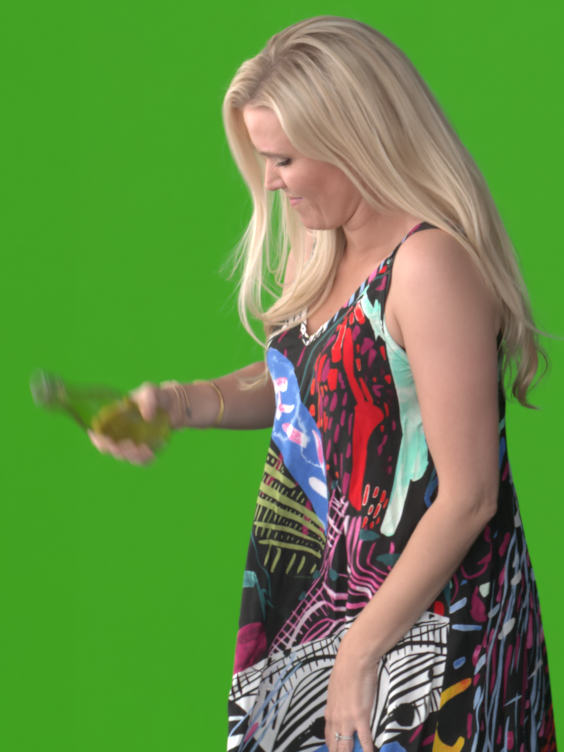}
         \vspace{-1.5em}
         \caption{Green background.}
         \label{fig:green-TMTM}
     \end{subfigure} \\\vspace{0.5em}
     \begin{subfigure}[t]{0.495\columnwidth}
         \centering
         \includegraphics[width=\textwidth]{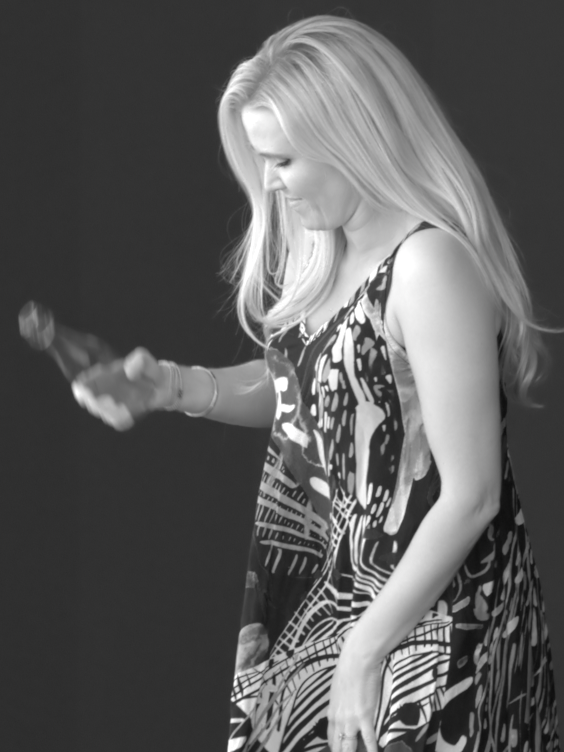}
         \vspace{-1.5em}
         \caption{Blue background (R channel).}
         \label{fig:blue-TMTM-red}
     \end{subfigure}
     \begin{subfigure}[t]{0.495\columnwidth}
         \centering
         \includegraphics[width=\textwidth]{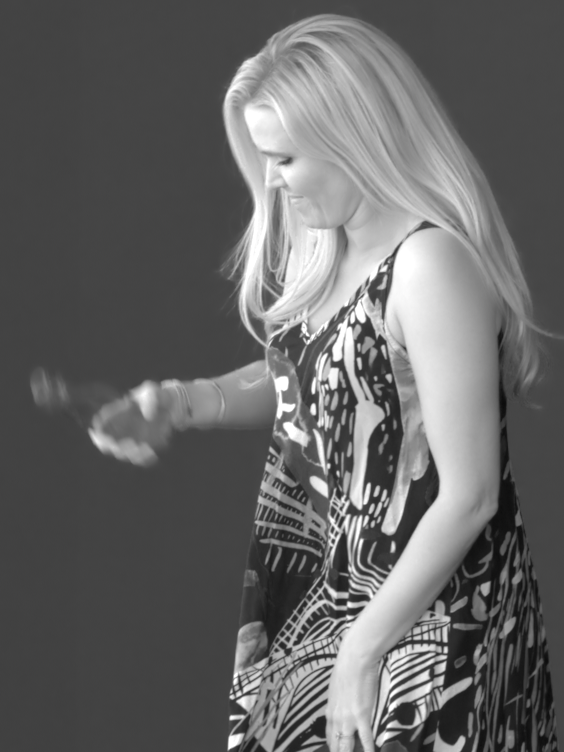}
         \vspace{-1.5em}
         \caption{Green background (R channel).}
         \label{fig:green-TMTM-red}
     \end{subfigure} \\\vspace{0.5em}
     \begin{subfigure}[t]{0.495\columnwidth}
         \centering
         \includegraphics[width=\textwidth]{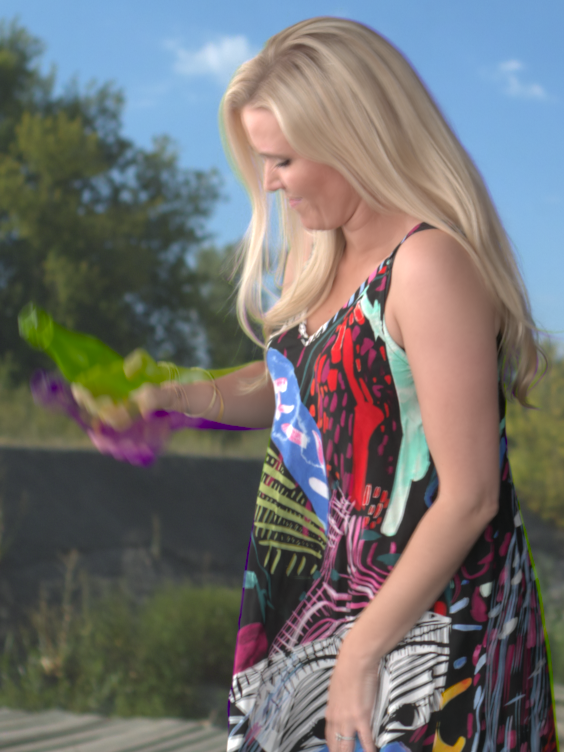}
         \vspace{-1.5em}
         \caption{Composite sans optical flow.}
     \end{subfigure}
     \begin{subfigure}[t]{0.495\columnwidth}
         \centering
         \includegraphics[width=\textwidth]{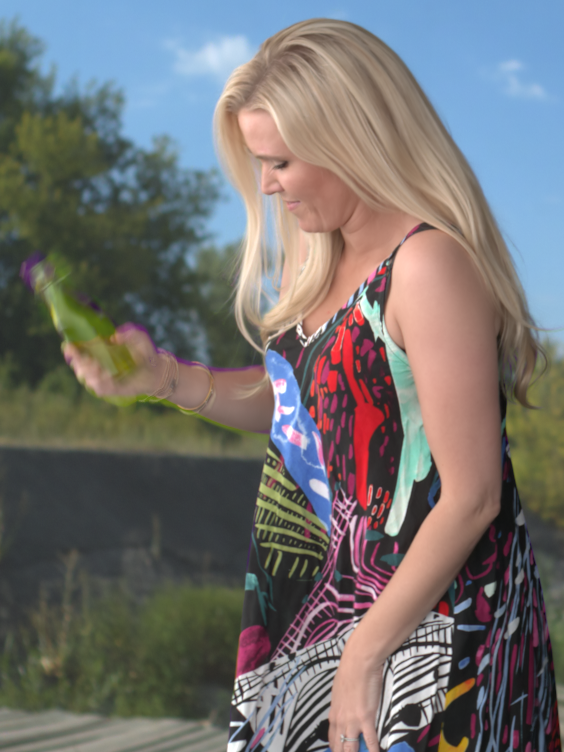}
         \vspace{-1.5em}
         \caption{Composite with optical flow.}
         \label{fig:TMTM-flow}
     \end{subfigure}
     \vspace{-0.5em}
        \caption{For Time-Multiplexed Triangulation Matting, we capture a white-lit subject behind alternating differently-colored frames (a, b). The red channels of the frames (c, d) are similarly lit and can be used for computing optical flow between the frames. We reconstruct the foreground without optical flow (e) and with optical flow (f).}
        \label{fig:TMTM}
\end{figure}

\citet{Smith:1996:Blue} proposed triangulation matting, where the alpha channel of a foreground subject is derived by seeing the foreground in front of two differently colored background images.  The technique was proposed for static scenes, but in this section we apply the technique to dynamic scenes using our time-multiplexing setup.  We keep static white LED lighting static on the actor and alternate the background LED panels in the camera frustum between green and blue, as in \Cref{fig:blue-TMTM,fig:green-TMTM}.  We note that after color matrixing to eliminate crosstalk, the red channel shows the actor lit by red light against a dark background in both frames, as seen in \Cref{fig:blue-TMTM-red,fig:green-TMTM-red}.  We thus can perform a more robust solution to the optical flow between the red channels of consecutive green and blue background frames, which are exposed just 1/48th of a second apart instead of 1/24th of a second.  Displacing the blue background frame to the position of the previous green background frame supplies the imagery needed for triangulation matting.  However, even with this improved optical flow technique, some temporal misalignment can remain, as in \Cref{fig:TMTM-flow}.

\section{Results and Discussion}

\subsection{Basic Magenta Green Screen Results}

\Cref{fig:teaser} shows the main steps of the Magenta Green Screen process.  \Cref{fig:teaser-mg} is a frame of a clip after the crosstalk elimination of \Cref{sec:crosstalk}.  \Cref{fig:teaser-alpha} shows the matte derived by dividing the green channel by its appearance in the clean plate and inverting.  \Cref{fig:teaser-naive} shows a naively colorized foreground element, where the green channel is replaced with a simple linear combination of the red and blue channels.  The bounce light subtraction, as in \Cref{sec:bouncelight}, has been applied to achieve a black background.  \Cref{fig:teaser-colorized} shows the foreground element colorized with machine learning, as described in \Cref{sec:colorize}, based on a color reference performance under white RGB light.  \Cref{fig:teaser-comp} shows the colorized foreground element composited onto a background image, exhibiting good matte edges and transparency for the bottles.  \Cref{fig:teaser-gt} shows a real ground truth comparison image, where the actors were illuminated by white RGB light and the background image was displayed on the LED panels as an in-camera visual effect.  Aside from slightly different actor poses, the composited image and real ground truth image are nearly indistinguishable, with believable matte edges and accurate image colorization.

The sequence in \Cref{fig:teaser} is shown in motion in the accompanying video, with the actors shaking the semitransparent bottles to generate varying degrees of motion blur in the footage, showing believable alpha transparency throughout.  

\begin{figure*}
    \centering
    \includegraphics[width=0.99\textwidth]{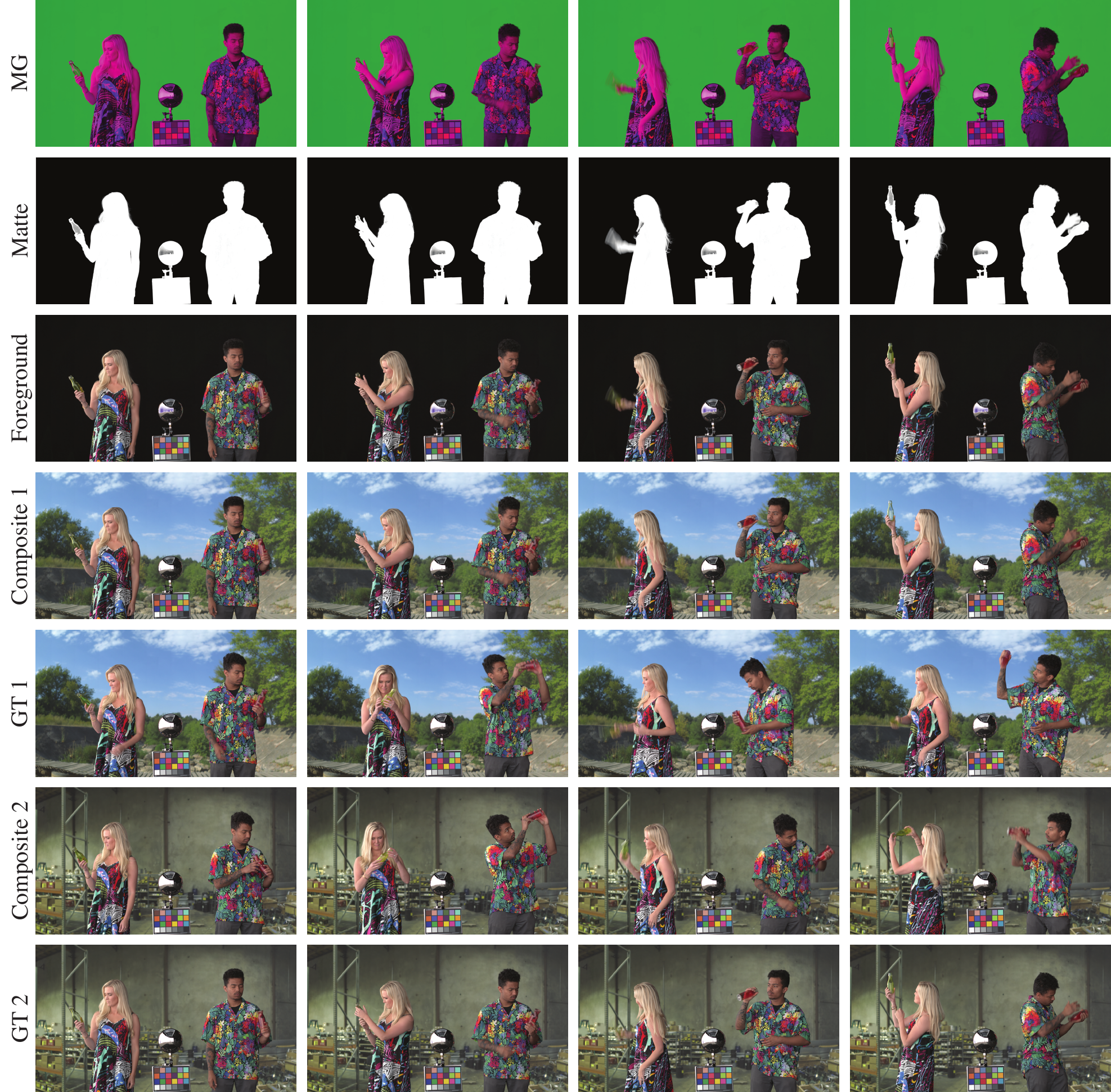}
    \vspace{-0.5em}
    \caption{Four frames of the performance sequence from the accompanying video, all processed and composited onto two different backgrounds using the basic Magenta Green Screen process. \emph{GT1} and \emph{GT2} show ground truth comparisons to the composited shots \emph{Composite 1} and \emph{Composite 2}.}
    \label{fig:results_grid}
\end{figure*}

\subsection{Which Color to Use for the Matte?}
\label{sec:yellowblue}

\begin{figure}
     \centering
     \begin{subfigure}[t]{0.48\columnwidth}
         \centering
         \includegraphics[width=\textwidth]{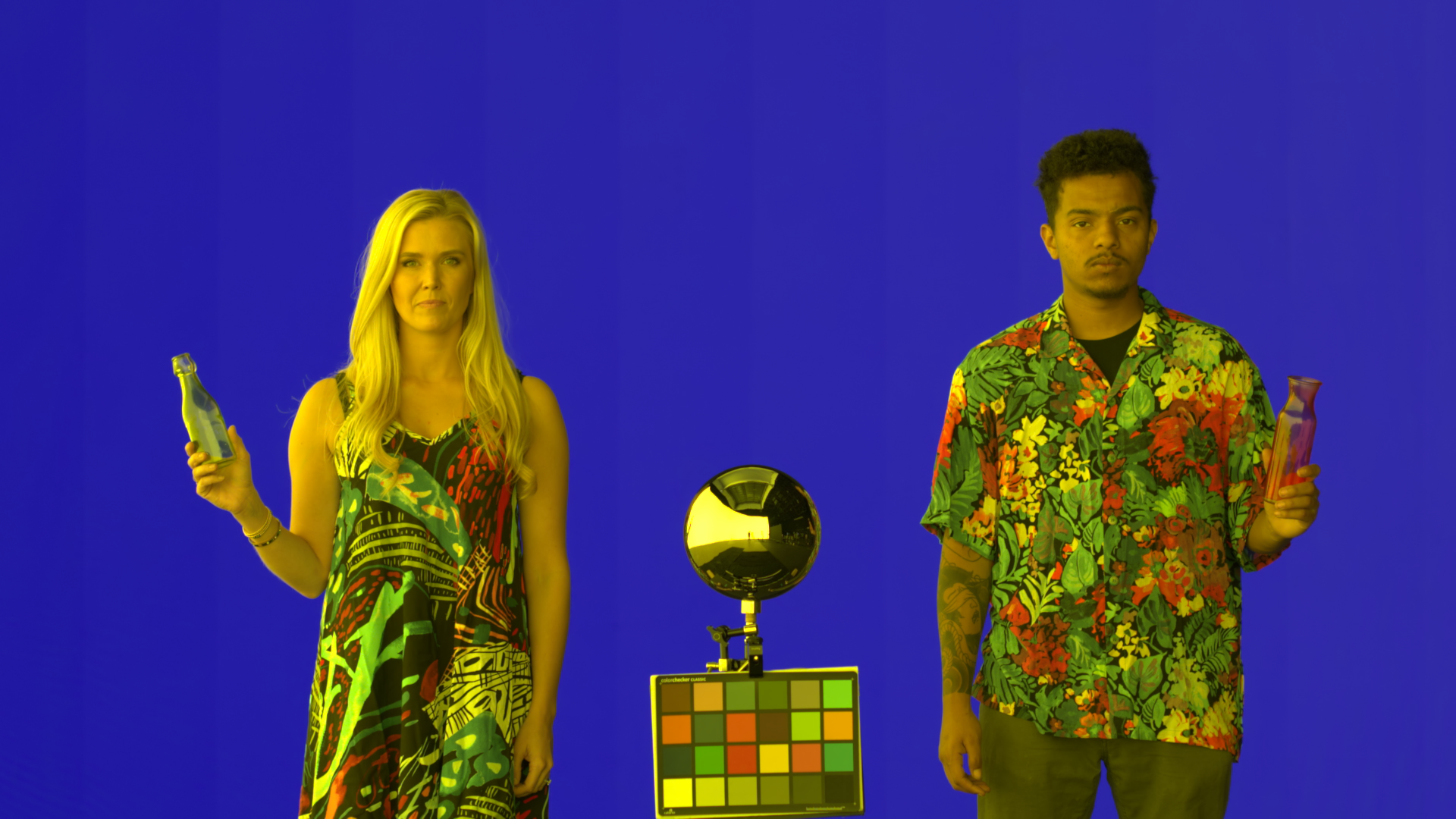}
         \caption{}
         \label{fig:yb}
     \end{subfigure}
     \hfill
     \begin{subfigure}[t]{0.48\columnwidth}
         \centering
         \includegraphics[width=\textwidth]{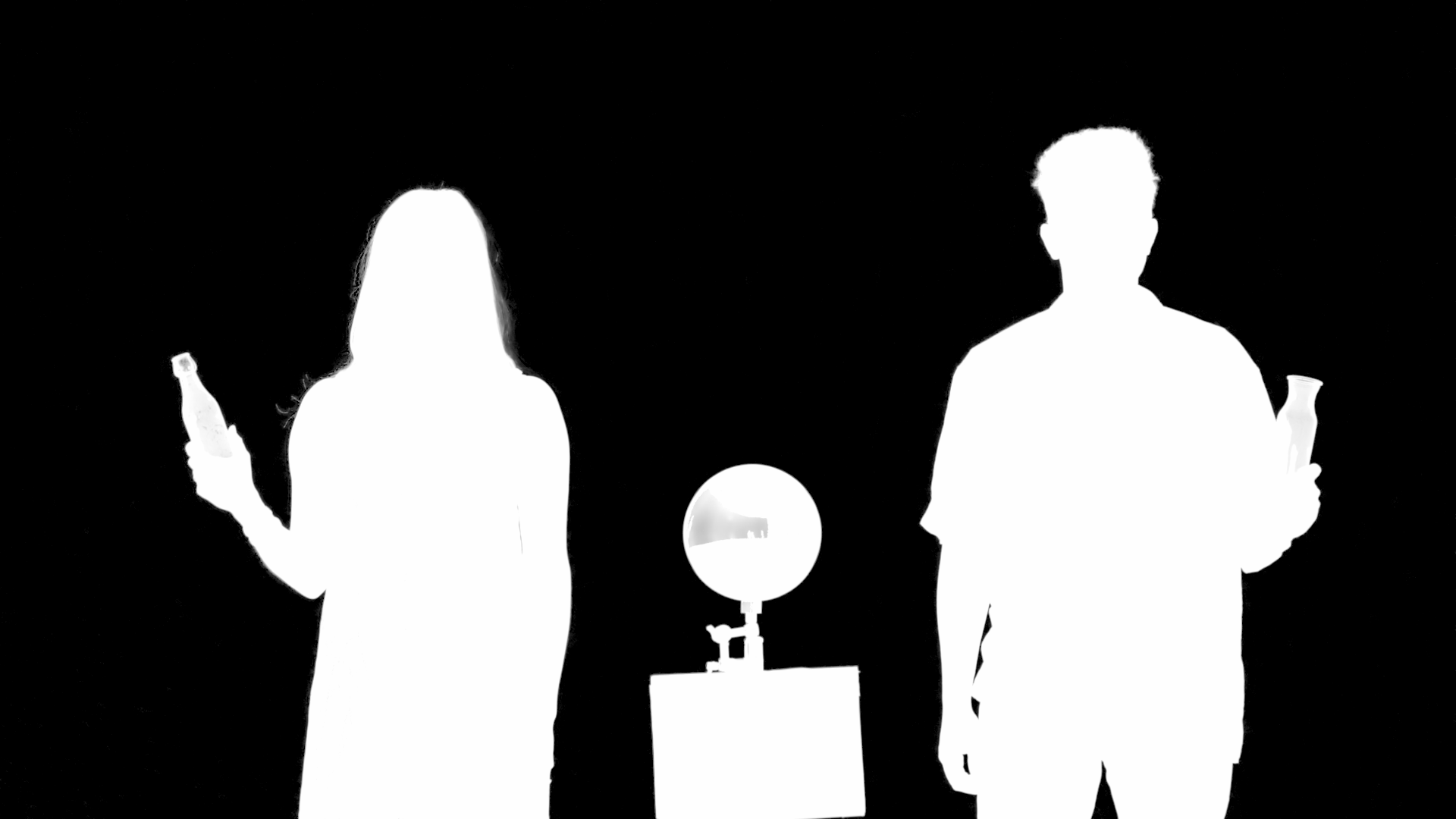}
         \caption{}
         \label{fig:yb-matte}
     \end{subfigure} \\
     \begin{subfigure}[t]{0.48\columnwidth}
         \centering
         \includegraphics[width=\textwidth]{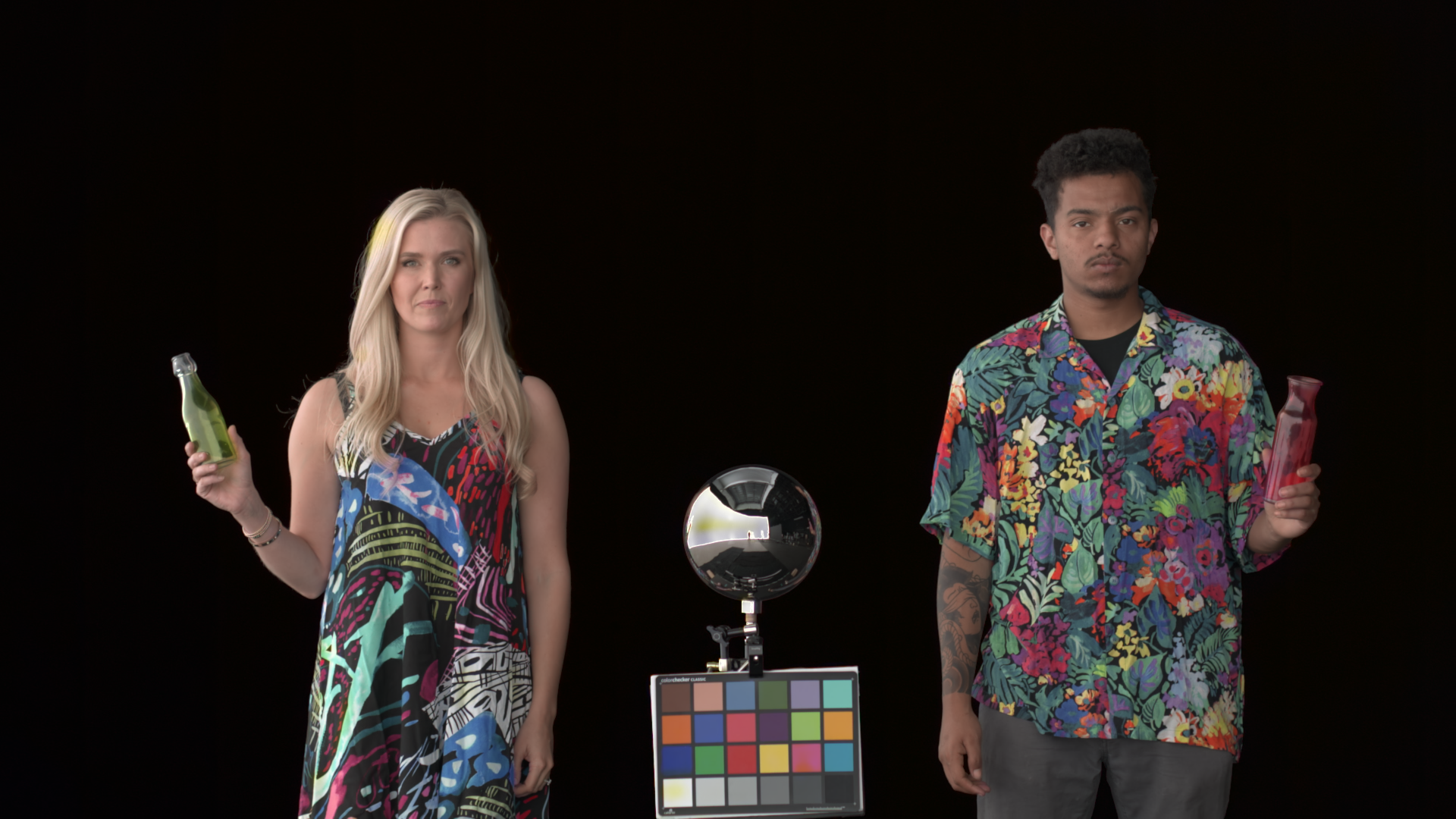}
         \caption{}
         \label{fig:yb-fg}
     \end{subfigure}
     \hfill 
     \begin{subfigure}[t]{0.48\columnwidth}
         \centering
         \includegraphics[width=\textwidth]{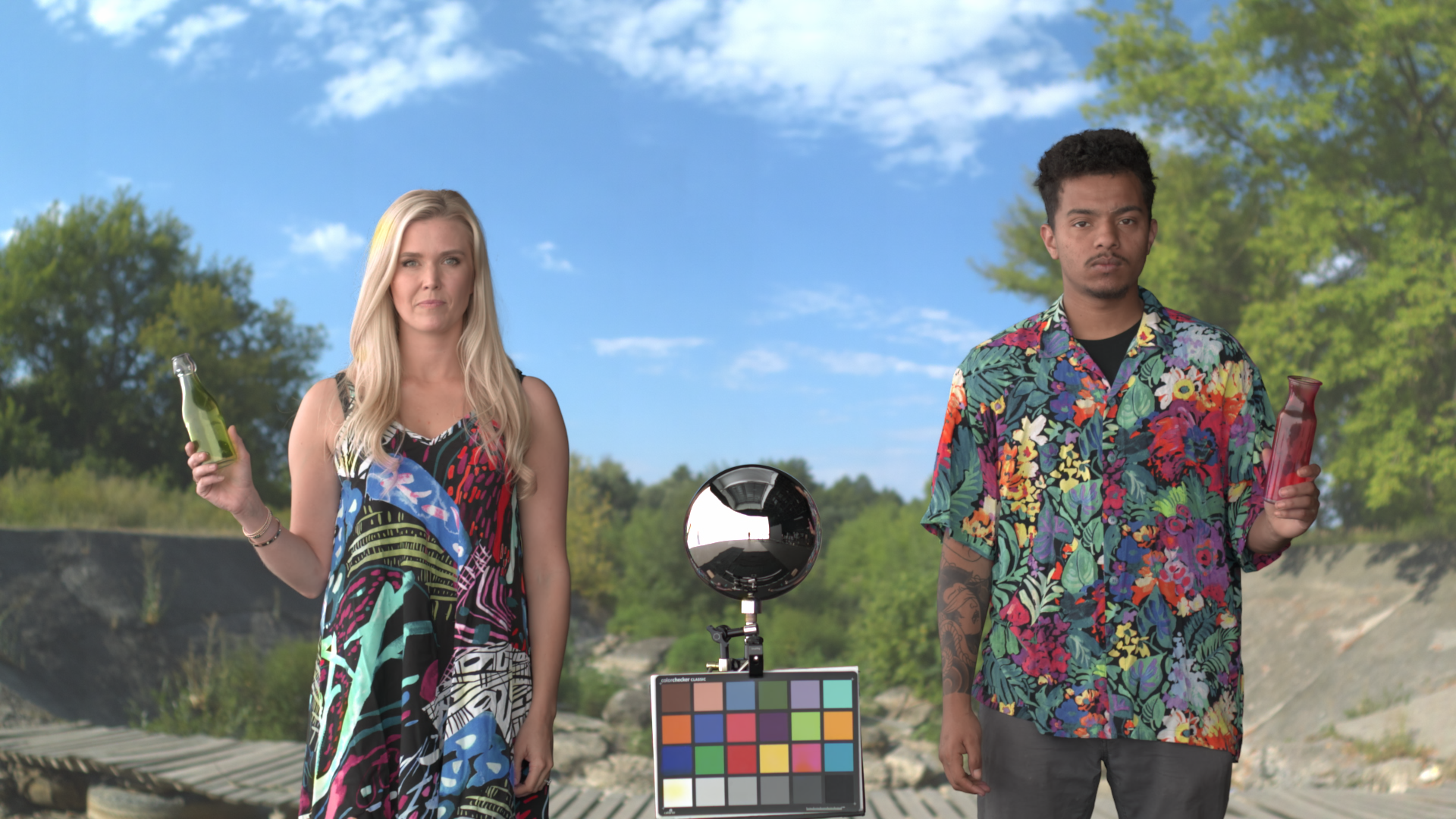}
         \caption{}
         \label{fig:yb-comp}
     \end{subfigure}
     \vspace{-1em}
        \caption{Yellow Blue Screen variant.  Actors are lit by red and green LED channels in front of a blue background (a). A matte is derived from the Blue channel (b). We colorize the foreground element (c) and produce a composite (d).}
        \label{fig:yellow-blue}
\end{figure}

We can alternatively choose to use the red or blue channel for recording the matte instead of green: this would result in Yellow Blue Screen matting or Cyan Red Screen matting.  We chose to use green for matting, since green screen is the most common traditional matting process and records the matte with the highest resolution channel, as Bayer pattern sensors have twice as many green pixels as they do red or blue.  We also imagined that inferring a green channel prompted by the red and blue channels might the easiest colorization process, since the algorithm needs to infer a color channel which is spectrally between two observed channels, rather than outside the spectral area which has been recorded.  To test this theory, we implemented the Yellow Blue Screen process as in \Cref{fig:yellow-blue}, training a different colorization process to predict the actors' blue channel from the red and green channels and the same RGB-white-lit reference sequence.  This technique also worked well, and the yellow light was somewhat more pleasing to look at than the magenta.  One artifact occurred on the side of the blonde hair, which appeared too yellow in some frames.  However, we believe this could be due to image sensor saturation in this area, since the blonde hair reflected more yellow light than it did magenta, and we left the exposures the same.

\subsection{Recovering a Full-Color Alpha Channel}

\begin{figure}
    \vspace{0.5em}
     \centering
     \begin{subfigure}[t]{0.48\columnwidth}
         \centering
         \includegraphics[width=\textwidth]{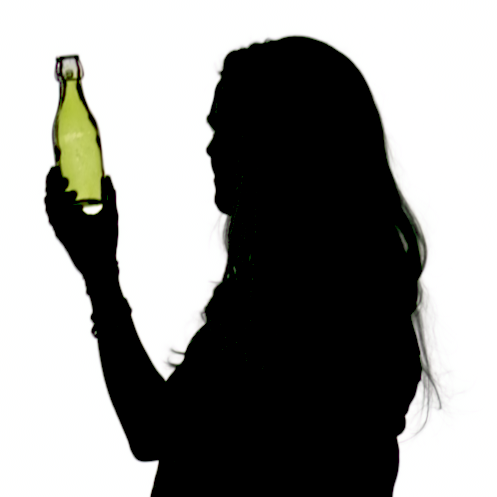}
         \caption{}
         \label{fig:fullcolorholdout}
     \end{subfigure}
     \hfill 
     \begin{subfigure}[t]{0.48\columnwidth}
         \centering
         \includegraphics[width=\textwidth]{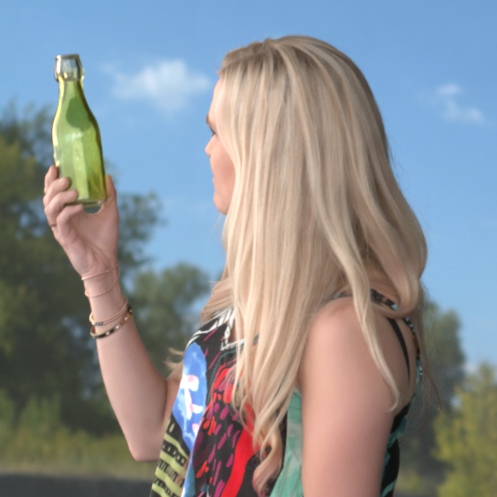}
         \caption{}
         \label{fig:fullcolorcomposite}
     \end{subfigure}
     \begin{subfigure}[t]{0.48\columnwidth}
         \centering
         \includegraphics[width=\textwidth]{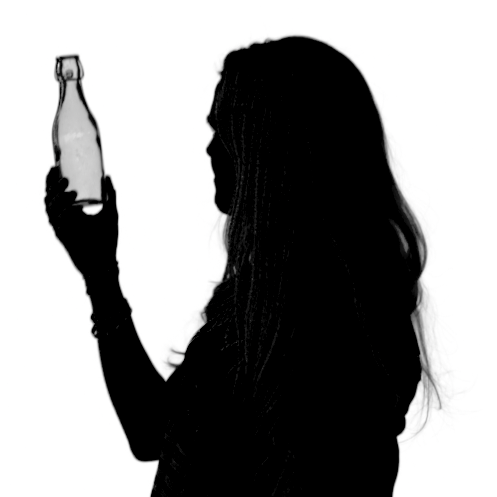}
         \caption{}
         \label{fig:monochromecolorholdout}
     \end{subfigure}
     \hfill 
     \begin{subfigure}[t]{0.48\columnwidth}
         \centering
         \includegraphics[width=\textwidth]{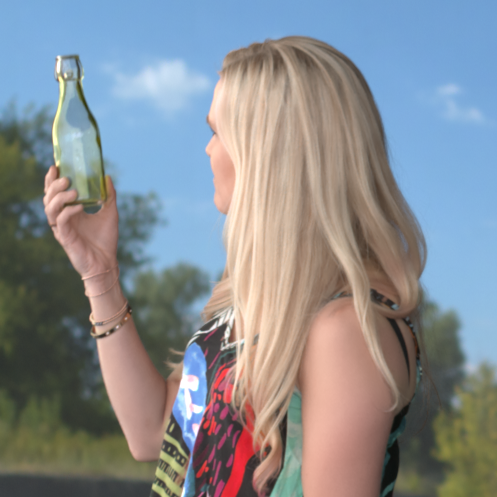}
         \caption{}
         \label{fig:monochromecomposite}
     \end{subfigure}
     \vspace{-1em}
        \caption{A full color matte (a) obtained by colorizing the monochrome holdout matte (c) from the green channel from Magenta Green Screen, showing colorful transparency of the bottles.  A composite (b) made using the full-color matte yields better color rendition of the bottle than done using the monochrome matte (d).}
        \label{fig:fullcolormatte}
\end{figure}

The alpha channel recovered from Magenta Green Screen is monochromatic, making the usual assumption that a foreground element's transparency in the green channel is the same for its red and blue channels.  If parts of the scene exhibit colorful transparency, such as the red and green glass bottles in \Cref{fig:teaser}, then a monochrome matte would have the objects transmit incorrectly neutral light, as in \Cref{fig:teaser-alpha}.  In this case, we can colorize the matte image from a reference recording of the actors performing while silhouetted in front of a white background.  Although this requires synthesizing two channels from one, and even though there is much less visual detail in the silhouetted imagery, our colorization framework is able to colorize the holdout matte image as well as in \Cref{fig:fullcolorholdout}.  To aid our model in matte colorization, we provide RGB channels of the frame prior to the input monochrome matte as additional signal---this training data can be obtained by multiplexing the silhouetted lighting with white lighting over a black background.

A composited result from this process is shown in \Cref{fig:fullcolorcomposite} and in the accompanying video.  This yields a subtle improvement in the appearance of the bottles compared to the basic Magenta Green Screen technique of \Cref{fig:teaser}.

\subsection{Time-Multiplexed Results and Comparisons}

The time-multiplexing technique results \Cref{sec:time-multiplexing} are shown in \Cref{fig:multiplexed} for Time-Multiplexed Magenta Green Screen, \Cref{fig:classicTMM} for Classic Time-Multiplexed Matting, and \Cref{fig:TMTM} for Time-Multiplexed Triangulation Matting.  A sequence processed from each technique is included in the accompanying video.  In each case, the technique works well except when there is significant subject motion, and optical flow is relied upon to align the channels of the foreground element and/or the matte from neighboring frames recorded at 48fps.  The ML colorization technique to reconstruct the green channel of Time-Multiplexed Magenta Green Screen has the advantage that no optical flow is required to align channels, and it can be applied to Time-Multiplexed Magenta Green Screen footage just as it can be to non-Time-Multiplexed Magenta Green Screen footage.

\subsection{Comparison to Traditional Green Screen}

\begin{figure}
     \centering
     \begin{subfigure}[t]{0.48\columnwidth}
         \centering
         \includegraphics[width=\textwidth]{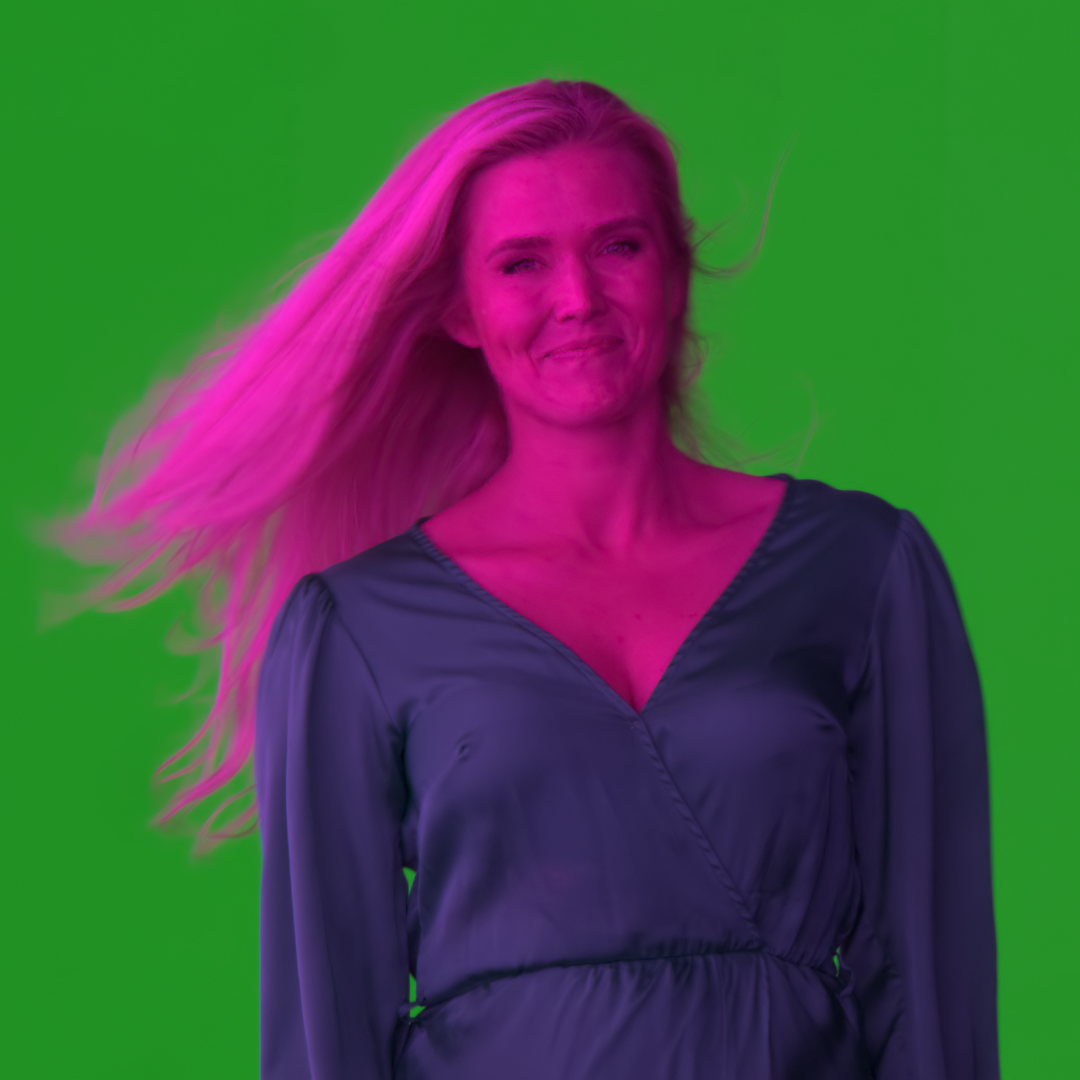}
         \caption{}
     \end{subfigure}
     \hfill
     \begin{subfigure}[t]{0.48\columnwidth}
         \centering
         \includegraphics[width=\textwidth]{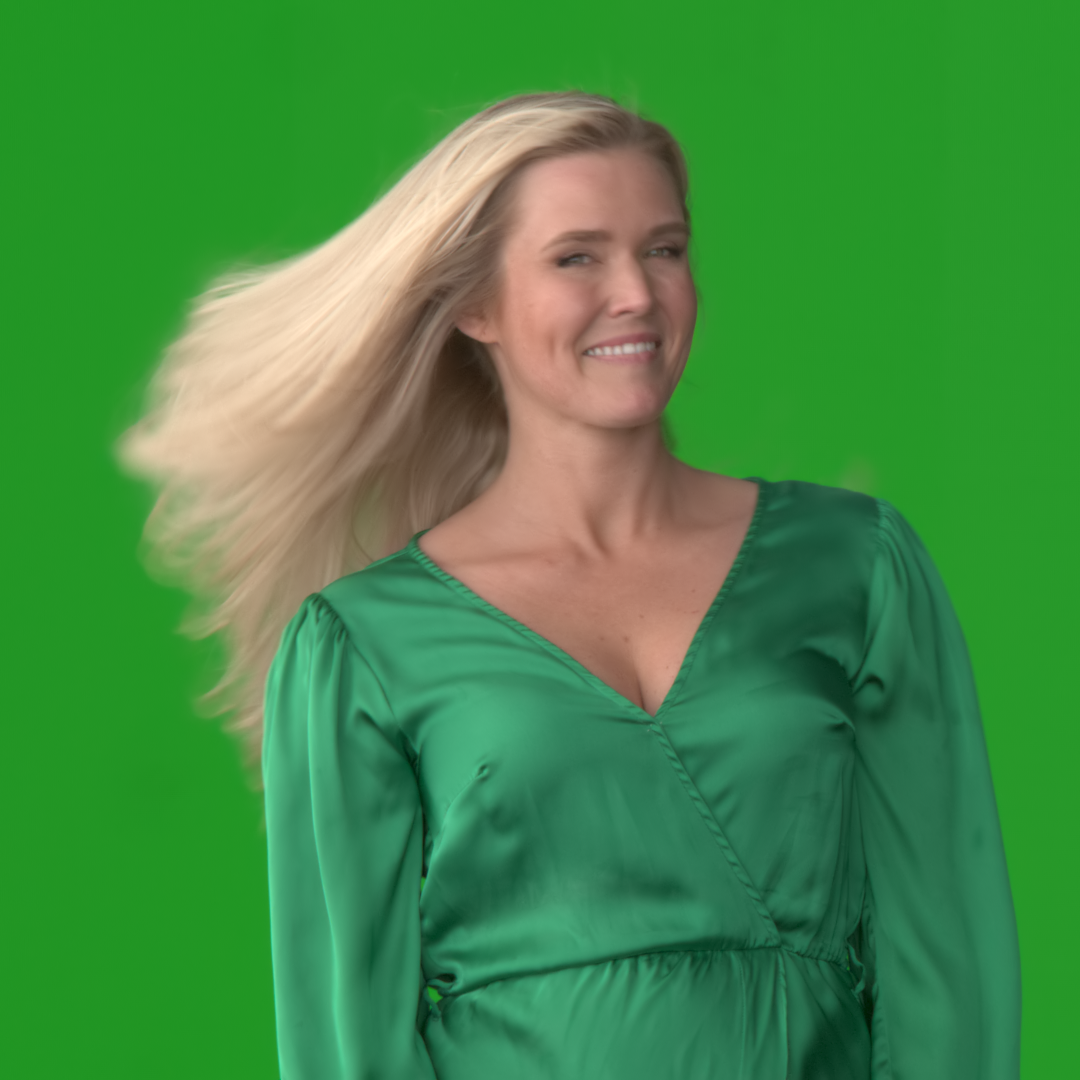}
         \caption{}
     \end{subfigure} \\
     \centering
     \begin{subfigure}[t]{0.48\columnwidth}
         \centering
         \includegraphics[width=\textwidth]{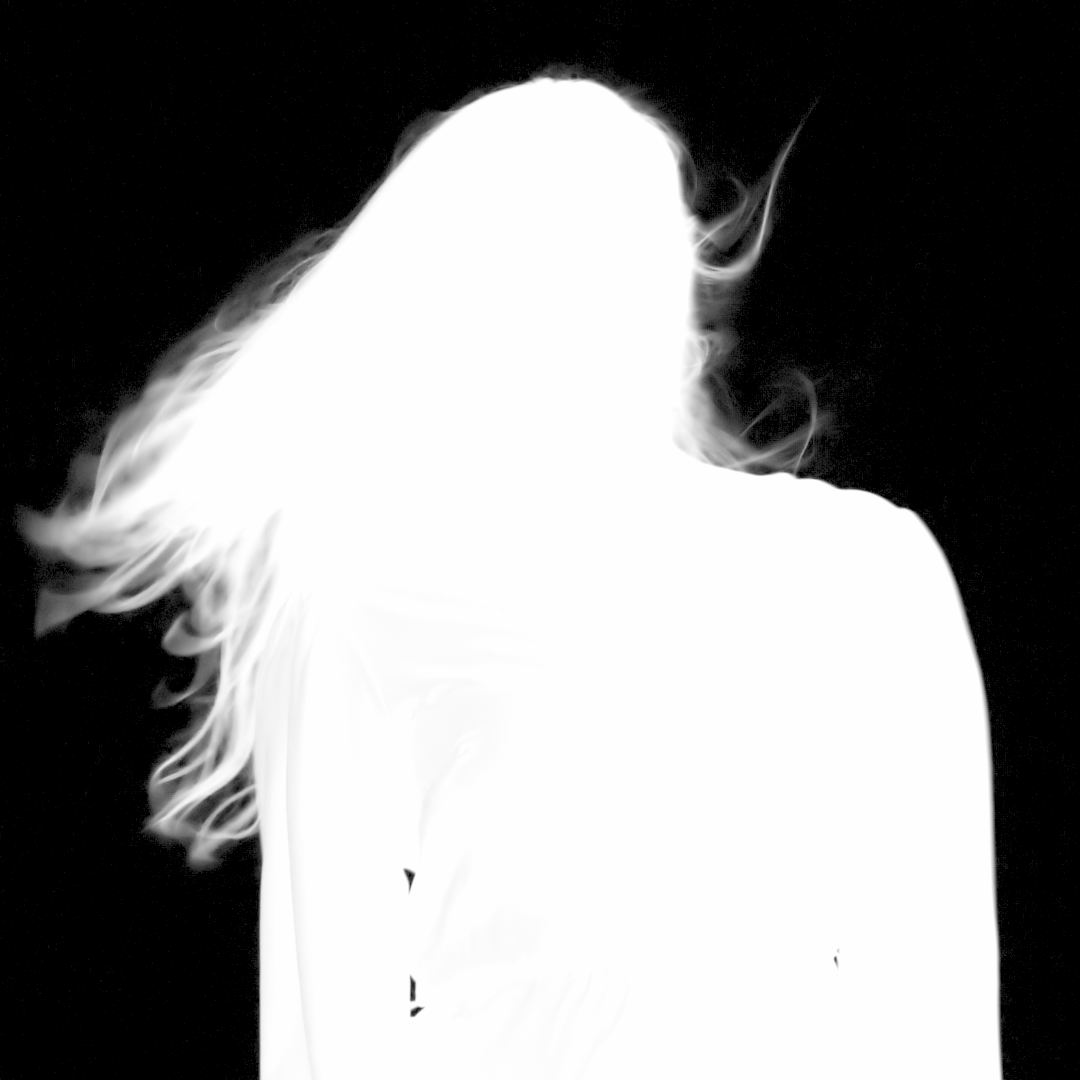}
         \caption{}
     \end{subfigure}
     \hfill
     \begin{subfigure}[t]{0.48\columnwidth}
         \centering
         \includegraphics[width=\textwidth]{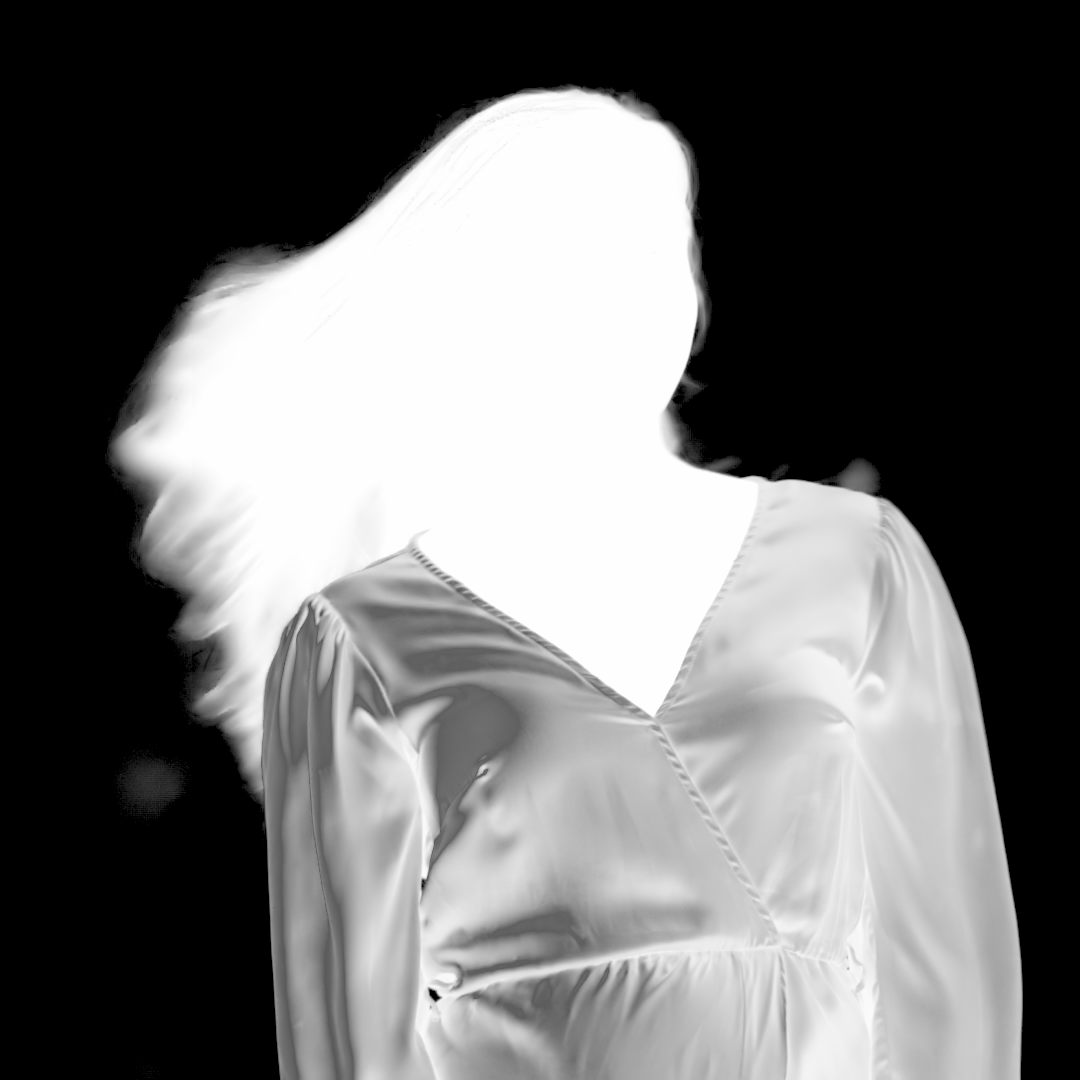}
         \caption{}
     \end{subfigure} \\
     \begin{subfigure}[t]{0.48\columnwidth}
         \centering
         \includegraphics[width=\textwidth]{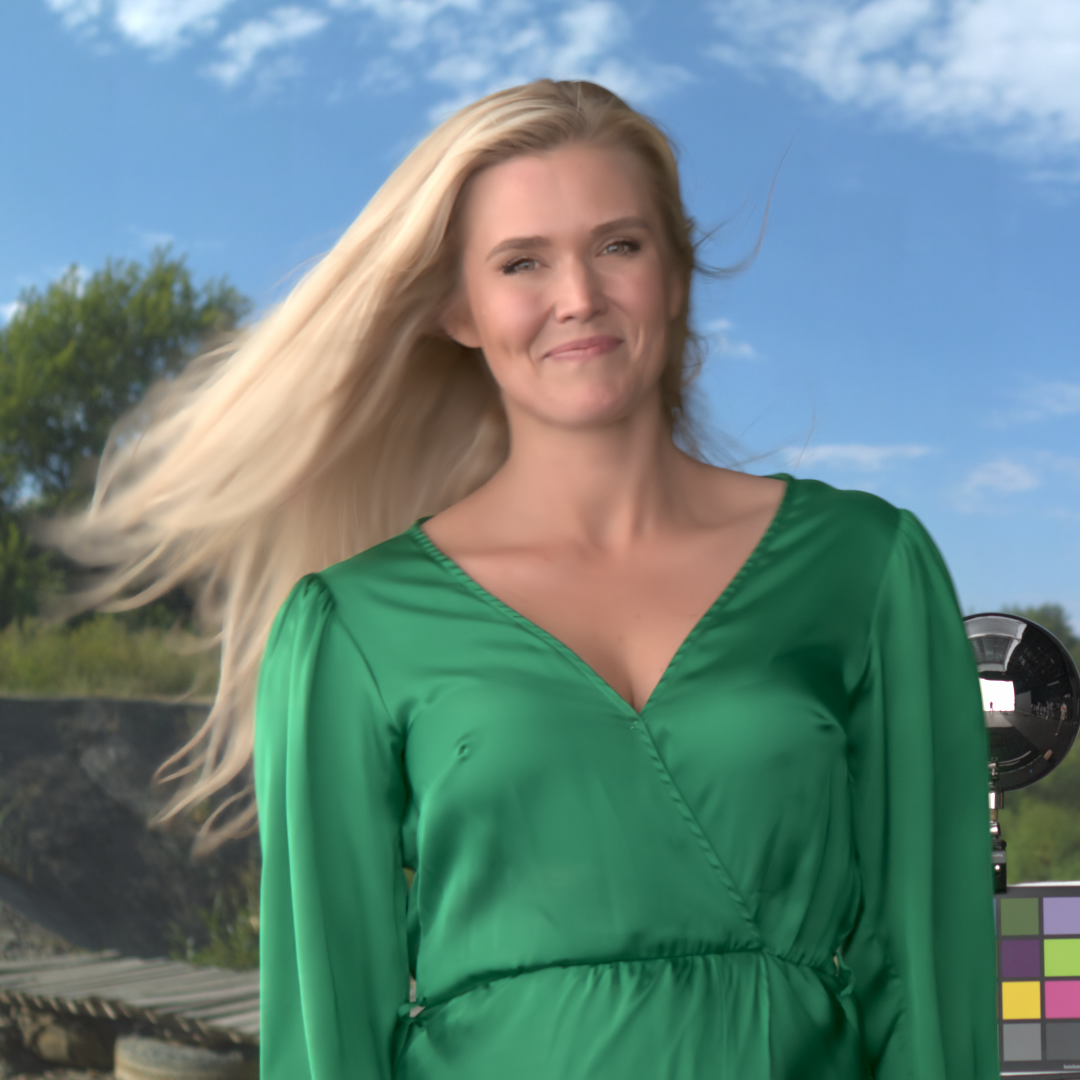}
         \caption{}
     \end{subfigure}
     \hfill 
     \begin{subfigure}[t]{0.48\columnwidth}
         \centering
         \includegraphics[width=\textwidth]{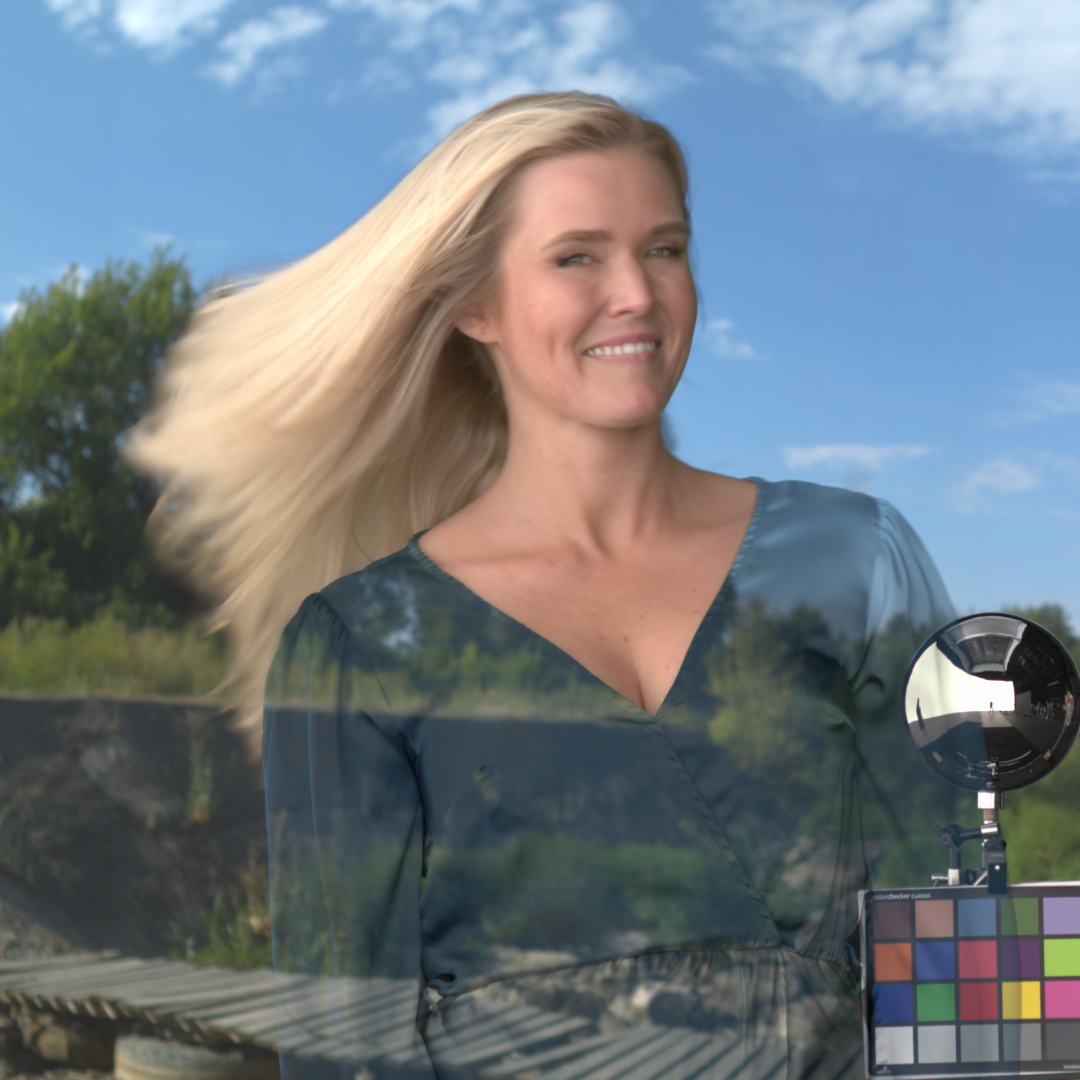}
         \caption{}
     \end{subfigure}
     \vspace{-1em}
        \caption{Comparison to traditional green screen. We record an actor lit with Magenta Green Screen lighting (a) as well as with white light against a traditional green screen (b). The matte generated from the Magenta Green Screen process (c) does not exhibit the artifacts of that generated from the traditional green screen using automated chroma keying (d). We show the corresponding Magenta Green Screen composite (e) and chroma keyed composite (f).}
        \label{fig:greenscreen}
\end{figure}

\Cref{fig:greenscreen} shows a matting comparison with a traditional green screen approach.  In this example, we use a subject wearing a green dress with long blonde hair blown by a fan.  The basic Magenta Green Screen approach yields a high-quality alpha channel and resulting composite, while an automated keying technique in Nuke applied to the footage struggles to key out the dress, rendering it significantly transparent; Magenta Green Screen also recovers somewhat more detail in the wispy hair.  While manual keying techniques could certainly succeed in keying this green screen shot, we are interested in an automated, reliable matting technique which makes no restrictions on what the actor wears or how they move.

\subsection{Discussion}

The main findings of this work are that the Magenta Green Screen approach appears to work well in generating high-quality alpha channels, and that our implementation of the colorization technique appears to be accurate, effective, and temporally stable.  Furthermore, our technique appears to outperform the matting that which can be obtained with either traditional chromakey green screen or time-multiplexed techniques which rely on optical flow to temporally align differently illuminated frames.  And while we did not compare to optical techniques to record a matte simultaneously with the foreground element (e.g., infrared matting and the sodium vapor process), we do not require a custom optical setup and do not need to align images from different sensors or cameras.
\section{Future Work}

Our method suggests a number of avenues for future work.  One desirable improvement would be to eliminate the need for recording a color reference clip of the actors in addition to the performance clip.  To this end, we might be able to develop a generalized green channel colorization model trained on large internet image collections, perhaps with some light augmentation of the model by reference photos of the actors from their costume tests.  Alternatively, we could try to use the green channels found in the alternate frames of the time-multiplexed Magenta Green Screen technique to train the colorization model.  This could be done by aligning the green channels to the preceding red and blue channels with optical flow, which perhaps would not need to be as accurate as when using the flow to align the frames for direct compositing.

Finally, it would be of use to implement a real-time version of the technique for on-set visualization and real-time in-camera visual effects.  While the colorization model training time is currently much too slow for this, one could imagine using the naive colorization technique while on set, and improving the colorization using the neural network training in post-production.
\section{Conclusion}

In this work, we have presented Magenta Green Screen, a new technique for alpha matte composting.  We use lighting conditions which are straightforward to produce on a modern LED virtual production stage combined with a data-driven machine learning algorithm, resulting in an approach that could make it easier to add or change the background environment behind actors more easily than current techniques allow.  A main finding of this work is that it can be very effective to train a network to restore one of the color channels to an RGB image, making it possible to use one of the channels to record an accurate alpha channel during the performance.  Another finding is that this image colorization process can produce fewer visual artifacts than time-multiplexed matting techniques which employ optical flow.  We look forward to applying these techniques to virtual production projects where being able to change the background in post-production is an important requirement.  Furthermore, the accurately measured alpha channels which can be obtained with our technique should be able to function as high-quality training data for natural image matting algorithms.
\balance

\begin{acks}
The authors wish to thank Eyeline Studios, Stephan Trojansky, Connie Siu, Kevin Izumi, and Lukas Lepikovsky for their support capturing data for this project, Lianette Albaner, Daniele Morkel, and Naz for acting in front of camera, and Kris Prygrocki for his assistance with the global shutter camera.
\end{acks}

\bibliographystyle{ACM-Reference-Format}
\bibliography{magenta_green}

\end{document}